\documentclass[11pt]{article}
\usepackage{amsmath}
\usepackage{graphicx}
\usepackage{lscape}
\usepackage{color}
\usepackage{orcidlink}
\usepackage{hyperref}
\hypersetup{colorlinks=true, linkcolor=blue, citecolor=blue, urlcolor=blue}
\usepackage{subcaption}
\RequirePackage[numbers,sort&compress]{natbib}
\begin{document}
\baselineskip=18pt

\begin{center}
{\large{\bf Morris-Thorne-type wormholes with global monopole charge and the energy conditions }}
\end{center}

\vspace{0.3cm}

\begin{center}
{\bf Jaydeep Goswami}\footnote{\bf email: jdpgsm98@gmail.com}, {\bf Hafizur Rahman}, {\bf Rimi Sikdar}, {\bf Rina Parvin} and {\bf Faizuddin Ahmed\orcidlink{0000-0003-2196-9622}}\footnote{\bf email: faizuddinahmed15@gmail.com}\\
\vspace{0.2cm}
{\it Department of Physics, University of Science \& Technology Meghalaya, Ri-Bhoi, Meghalaya, 793101, India}\\

\end{center}

\vspace{0.3cm}

\begin{abstract}
In this paper, we investigate Morris-Thorne-type wormholes with global monopole charge using various shape function forms known in the literature. We solve the Einstein field equations incorporating an anisotropic energy-momentum tensor and obtain different physical quantities associated with the matter- content. A crucial aspect of this study is the non-exotic matter distribution, examined through the evaluation of energy conditions, and exploring how different shape functions impact these conditions. Additionally, the anisotropy parameter is calculated to quantify the extent of attractive or repulsive behavior. Our study demonstrates that for different types of shape function forms, the energy conditions are influenced by the global monopole parameter. Our findings provide valuable insights for further theoretical explorations of these fascinating hypothetical structures in the realm of general relativity and beyond. 
\end{abstract}

\vspace{0.1cm}


\vspace{0.1cm}

{\bf PACS Number(s)}: 04.50.Kd; 04.20Gz; 04.20.Jb; 14.80.Hv

\section{Introduction}

Wormholes are hypothetical structures in space-time that connects two distinct universes, called inter-universe wormholes or two different regions of the same space-time in the same universe; called intra-universe wormholes. The difference between these two classes of wormhole ascends only at the global level of geometry and global level of topology. These tube-like structures are considered to be asymptotically flat at both ends. Wormholes can be either static, with a constant throat radius, or non-static, with a variable throat radius. The throat of a wormhole is the narrowest part of the tube connecting the two mouths, potentially linking two universes. Near the throat of the wormholes local physics is oblivious to issues of intra-universal or inter universal travel. Flamm \cite{aa1} first studied wormhole type solutions. Einstein and Rosen carried out the study of these geometrical objects and obtain a solution involving event horizons which is known as Einstein-Rossen bridge \cite{aa2} .The word wormhole was first introduced by Misner and Wheeler \cite{aa3}. They showed that wormholes cannot be traversable for standard matter due to its instability. Ellis \cite{aa4} and Bronnikov \cite{aa5} first assumed an example of wormhole and the corresponding solution nowadays known as Ellis-Bronnikov wormhole. Later on, this wormhole was found to be unstable \cite{aa6,aa7}. The interest in the study of wormholes started with the work of Morris and Thorne \cite{aa8}. They provided some conditions for a wormhole to be traversable. They showed that the presence of exotic matter, matter different from the normal matter, is necessary for the existence of traversable wormholes. In Ref. \cite{aa9}, it was showed that the wormhole structure in space-time, developed for inter-stellar travel, can be transformed into time-machine.

Wormholes, also known as astrophysical compact objects, are distinct, non-trivial topological structures that are free from singularities and event horizons \cite{cc1}. These structures have been studied using various approaches, including the application of specific equations of state, constraining fluid parameters, and solving for metric elements. Geometrically, wormholes are often conceptualized as potential mechanisms for warp drives, time travel, and rapid interstellar travel. Within the framework of general relativity (GR), static and spherically symmetric Lorentzian wormhole solutions are a subject of extensive research. A key aspect in understanding the properties and behavior of traversable wormholes is the shape function, which mathematically describes the spatial geometry of the wormhole, particularly the relationship between the radius of the throat and the radial coordinate \cite{cc2}.

Exotic matter involves the violation of the null energy condition. At the throat of a wormhole, the stress energy distribution is notably abnormal. When converging radial light rays enter a wormhole, they reach the throat and lose null convergence condition, becoming parallel at the throat before diverging on the other side. Consequently, the null convergence condition and thus the null energy condition, is not satisfied at the throat in general relativity \cite{aa10}. This leads to the matter near the wormhole throat not satisfying the null energy condition and negative energy density may also be observed in this region. As a result, other energy conditions are violated as well \cite{aa11,aa12}.

However, in modified theories of gravity, violating energy conditions is not required because the function in the gravitational action differs from that in General Relativity, leading to completely different field equations. Among the various alternative theories developed in the literature, the $f(R)$ theory of gravity is one that generalized Einstein’s theory of general relativity. It incorporates an arbitrary function $f(R)$ of the Ricci scalar $R$ into the gravitational action, instead of a function equal to $R$. This theory has been studied in various aspects \cite{aa15,aa21,aa22,aa23,aa24,aa25,aa26,aa27,aa28,aa29,aa30,aa31,aa32,aa33,aa34}.

The study of wormhole solutions has become a significant area of research, attracting the attention of various cosmologists who have effectively utilized $f(R)$ theory. Lemos {\it et al.} \cite{aa35} reviewed the existing literature on wormholes, analysed the Morris-Thorne metric with the inclusion of a cosmological constant, and examined the properties of traversable wormholes. Rahaman {\it et al.} \cite{aa36} treated the cosmological constant as a function of the radial coordinate, derived wormhole solutions, and established a condition for the averaged null energy condition to be small. Lobo {\it et al.} \cite{aa37} explored the existence of wormholes within the $f(R)$ theory of gravity. Garcia {\it et al.} \cite{aa38} studied wormholes by assuming a coupling between a function of scalar curvature and the Lagrangian density of matter, using linear and non-minimal couplings to obtain an exact solution. Duplessis {\it et al.} \cite{aa39} applied scale-free $R^2$ gravity, discovering exotic solutions for traversable wormholes and black holes. Bahamonde {\it et al.} \cite{aa40} analysed wormhole solutions in modified $f(R)$-gravity, employing equations of state for matter and radiation-filled universes to reveal a wormhole structure asymptotically approaching an FLRW universe. Moradpour \cite{aa41} investigated traversable wormholes in general relativity and Lyra manifold, and found some asymptotically flat solutions. Zubair {\it et al.} \cite{aa42} utilized $f(R,\varphi)$ gravity to study symmetric, spherical, and static wormhole solutions, analyzing energy conditions for various fluid types and identifying constraints for weak and null energy conditions, suggesting realistic wormhole structures. Shaikh \cite{aa43} found wormhole solutions in Eddington-inspired Born-Infeld gravity with non-exotic matter satisfying energy conditions. Barros {\it et al.} \cite{aa44} demonstrated that three-form fields could support wormhole geometries, validating weak and null energy conditions. Peter \cite{aa45} used $f(R)$-gravity to investigate traversable wormholes, deriving several solutions based on different shape function and showing that non-commutative geometry combined with $f(R)$-gravity leads to null energy condition violation. Maurya {\it et al.} used matter coupling gravity formalism \cite{cc3} with observational data to look for potential wormhole solutions. In addition, investigations of wormhole models in other modified gravity theories, such as $f(\mathrm{Q},\mathrm{T})$-gravity, $f(R,\mathrm{T})$-gravity, where $\mathrm{T}$ is the trace of the stress-energy tensor, $f(R,\mathcal{L}_m)$, $f(\mathrm{Q})$-gravity, and Rastall gravity etc. have been reported in the literature (see, for examples, Refs. \cite{bb1,bb2,bb3,bb4,bb5,bb6,bb7,bb8,bb9,bb10,bb11,bb12,bb13,bb14,bb15,bb16,bb17,bb18,bb19}). The possibility of construction of a traversable wormhole on the Randall–Sundrum braneworld, with non-exotic matter employing the Kuchowicz potential, has also been studied \cite{cc4}.

Many cosmologists have significantly contributed to the study of wormholes. Over the years, different scientists have proposed various forms of wormholes, such as the Kerr wormhole \cite{aa46}, the Butcher wormhole \cite{aa47}, Rotating wormholes \cite{aa48,aa49,aa50}, and defect wormholes \cite{aa51,aa52,aa53,aa54,aa5,aa55}. These wormholes were represented using different metric tensors. Despite the establishment of the theory, maintaining the stability of wormholes remains a major challenge. They require a hypothetical substance called exotic matter, which has negative energy density, to stay open. 

Wormhole arises from a special solution to Einstein’s field equation \cite{aa11,aa12}. The Einstein field equations without cosmological constant $(\Lambda=0)$ are given by
\begin{equation}
    G_{\mu\nu}=R_{\mu\nu}-\frac{1}{2}\,R\,g_{\mu\nu}=T_{\mu\nu}\quad (\mu,\nu=0, 1, 2, 3), \label{a1}
\end{equation}
where $g_{\mu\nu}$ is the metric tensor with $g^{\mu\nu}$ its contravariant form, $R_{\mu\nu}$ is the Ricci tensor, $R=g^{\mu\nu}\,R_{\mu\nu}$ is the Ricci scalar, and $T_{\mu\nu}$ is the energy-momentum tensor.

Energy conditions are the mathematical constraints that dictate how matter and energy behave within space- time. In general relativity, all physically realizable energy-momentum tensors are expected to fulfill the energy conditions. The different energy conditions are as follows \cite{aa11,aa12}:

\begin{enumerate}
    \item {\bf The Null Energy Condition (NEC)}:
    For any null vector $k_{\mu}$, the energy-momentum tensor $T_{\mu\nu}$ satisfies the NEC provided
    \begin{equation}
        T_{\mu\nu}\,k^{\mu}\,k^{\nu} \geq 0,\quad\quad k^{\mu}\,k_{\mu}=0.\label{a2} 
    \end{equation}
    Or equivalently, we can write $\rho+p_i \geq 0\quad  \forall \quad i \in [1,2,3]$ for a fluid.

    \item {\bf The Weak Energy Condition (WEC)}:
    For any time-like vector field $U^{\mu}$, the energy-momentum tensor $T_{\mu\nu}$ satisfies the WEC provided
    \begin{equation}
        T_{\mu\nu}\,U^{\mu}\,U^{\nu} \geq 0,\quad\quad U^{\mu}\,U_{\mu}=-1.\label{a3}
    \end{equation}
    Or equivalently, we can write $\rho \geq 0$ and $\rho+p_i \geq 0\quad \forall\quad  i \in [1,2,3]$ for a fluid.

    \item {\bf The Strong Energy Condition (SEC)}:
    For any time-like vector field $U^{\mu}$, the energy-momentum tensor $T_{\mu\nu}$ satisfies the WEC provided
    \begin{equation}
        \left(T_{\mu\nu}-\frac{1}{2}\,g_{\mu\nu}\,T\right)\,U^{\mu}\,U^{\nu} \geq 0.\label{a4}
    \end{equation}
    Here, $T$ is trace of the energy-momentum tensor $T_{\mu\nu}$ given by $T=g_{\mu\nu}\,T^{\mu\nu}$. In other words, we can write $\rho+p_i+2\,p_j \geq 0\quad \forall\quad i,j \in [1,2,3]$ for a fluid.

    \item {\bf The Dominant Energy Condition (DEC)}:
    For any time-like vector field $U^{\mu}$, the energy-momentum tensor $T_{\mu\nu}$ satisfies the DEC provided
    \begin{equation}
        T_{\mu\nu}\,U^{\mu}\,U^{\nu} \geq 0\quad \mbox{and}\quad T_{\mu\nu}\,U^{\nu}<0.\label{a5}
    \end{equation}

    Or equivalently, $\rho-|p_i| \geq 0\quad \forall\quad i \in [1,2,3]$ for a fluid.
    
\end{enumerate}

In general relativity, anisotropy parameter refers to a quantity that measure the degree of anisotropy or directional dependence of a physical system or space-time geometry. It is defined by
\begin{equation}
    \Delta=p_t-p_r,\label{a6}
\end{equation}
where $p_r$ and $p_t$ are the radial and tangential pressures, respectively of a fluid. The space-time geometry is either attractive or repulsive if the anisotropy parameter $\Delta <0$\, or\, $\Delta>0$. For $\Delta=0$, the matter-content fluid is isotropic in nature. That means pressure is same in all spatial directions. 

The line element for a spherically symmetric and static four-dimensional traversable wormhole in ``Schwarzschild coordinate" $(x^0=t, x^1=r, x^2=\theta, x^3=\phi)$ is given by \cite{aa8,aa9}
\begin{equation}
    ds^2=-e^{2\,\Phi(r)}\,dt^2+\frac{dr^2}{\left(1-\frac{A(r)}{r}\right)}+r^2\,d\theta^2+r^2\,\sin^2\theta\,d\phi^2.\label{a7} 
\end{equation}

The radial coordinate $r$ has a range that increases from a minimum value at $r_0$, corresponding to the radius of the wormhole throat, to $\infty$, i.e., $r \in [r_0,\infty)$ and other coordinates are $-\infty < t < +\infty$, $0 \leq \theta < \pi$, and $0 \leq \phi \leq 2\,\pi$. Here $\Phi(r)$ is the red shift function and $A(r)$ is the shape function. The shape function $A(r)$ in line-element (\ref{a7}) must satisfy flare-out condition \cite{aa8,aa9,aa10}. At the wormhole throat, the shape function $A(r)$ must satisfy the condition
\begin{equation}
    A(r)|_{r=r_0}=r_0.\label{a8}
\end{equation}
Although the metric is singular at $r=r_0$, proper length/distance as an invariant quantity must be well-behaved, and therefore, the following integral must be real and regular out the throat \cite{aa8,aa9,aa10}
\begin{equation}
    \ell(r)=\pm\,\int^{r}_{r_0}\,\frac{dr}{\sqrt{1-A(r)/r}}.\label{a9}
\end{equation}
For the upper universe of wormhole, proper length $\ell(r)$ is positive, and for the lower universe, $\ell(r) < 0$.

The existence of wormhole solutions demands the satisfaction of the conditions: (i) $\frac{A(r)}{r} <1$ for $r>r_0$, (ii) the flaring out condition $A'(r=r_0) -1 \leq 0$, where superscript $(')$ denotes ordinary radial derivative w. r. t. $r$, (iii) asymptotically flat geometry, $\frac{A}{r} \to 0$ as $r \to \infty$, (iv) $\frac{A(r)-r\,A'(r)\,}{A^2(r)} >0$. For simplicity, the redshift function $\Phi(r)$ is assumed to be a constant so that there is no gravitational redshift. 
 
In this paper, we aim to investigate a topologically charged wormholes analogues to the Morris-Thorne-type wormholes space-time. We solve the Einstein field equations by choosing different shape functions and discuss the outcomes for anisotropic fluid as matter content. We demonstrate that the presence of the global monopole parameter controls the energy conditions for different shape functions form, ensuring they obey the energy conditions and thus remain free from exotic matter distribution. Moreover, we show that a few wormholes models are attractive and some other repulsive in nature. This paper is summary as follows: In section 2, we present a metric ansatz of topologically charged Morris-Thorne-type wormhole space-time and solve the Einstein field equations by choosing anisotropy as matter-energy content. Then, we choose various shape functions proposed in the literature and obtain the result and analyze them. In Section 3, we present our results and discussions.  Throughout the analysis, we choose the system of units where $c=1=G$.

\section{Wormhole models with global monopole charge and the energy conditions}

In Ref. \cite{aa57}, the author constructed a global monopole space-time represented by the following line-element
\begin{equation}
    ds^2=-\left(1-8\,\pi\,\mathrm{G}\,\eta^2-\frac{2\,M}{r}\right)\,dt^2+\frac{dr^2}{\left(1-8\,\pi\,\mathrm{G}\,\eta^2-\frac{2\,M}{r}\right)}+r^2\,(d\theta^2+\sin^2\theta\,d\phi^2),\label{Barriola}
\end{equation}
where $M$ is a constant of integration and in flat space $M \sim M_{core}$ ( $M_{core}$ is the mass of the monopole core), $\eta$ being the energy scale of symmetry breaking, and $\mathrm{G}$ is the Newton's gravitational constant. By neglecting the mass term and rescaling the variables $r$ and $t$ \cite{aa57}, one may rewrite a point-like global monopole metric as follows \cite{aa57,aa58,aa59}:
\begin{equation}
    ds^2=-dt^2+\frac{dr^2}{\alpha^2}+r^2\,(d\theta^2+\sin^2\theta\,d\phi^2),\label{Barriola2}
\end{equation}
where $\alpha^2=(1-8\,\pi\,\mathrm{G}\,\eta^2)$ with $\alpha$ is the global monopole charge\footnote{The topological charge, $Q$, whose origin is the GM model and
can be calculated following its definition in Ref \cite{AV}, $Q=\frac{1}{8\,\pi} \oint\,dS^{ij}\,|\phi|^{-3}\,\varepsilon_{abc}\,\phi^{a}\,\partial_{i}\phi^{b}\,\partial_{j}\phi^{c}$ and the matter source is given in Ref. \cite{aa57}.} that depends on the energy scale $\eta$. Noted that the presence of global monopole in space-time produces a solid deficit angle by an amount $\Delta\Omega=32\,\pi^2\,G\,\eta^2$. The physics associated with this point-like global monopole space-time has been discussed in details in Ref. \cite{aa58}.

Inspired by this, we introduce a metric anstaz that represents Morris-Thorne (MT)-type wormholes with global monopole charge. Therefore, line-element describing this MT-type wormhole with global monopole charge in ``Schwarzschild coordinate" $(x^0=t,\,\, x^1=r,\,\, x^2=\theta,\,\, x^3=\phi)$ is given by
\begin{equation}
    ds^2=-e^{2\,\Phi(r)}\,dt^2+\frac{dr^2}{\alpha^2\,\left(1-\frac{A(r)}{r}\right)}+r^2\,(d\theta^2+\sin^2\theta\,d\phi^2),\label{b1}   
\end{equation}

The presence of the global monopole charge $\alpha$ in a solution is referred to as a topologically charged space-time model. Numerous authors have studied topologically charged space-times within the framework of general relativity, as well as in the context of modified gravity theories proposed by various researchers. These includes lensing phenomena and gravitational field in a topologically charged Eddington-inspired Born–Infeld space-time \cite{pp1,pp2}, nonlinear $\sigma$-models minimally coupled to the Eddington-inspired Born-Infeld gravity \cite{pp3}, $f(R)$ global monopole space-time \cite{pp4,pp5}, gravitational lensing phenomena and thermodynamic properties of a black hole in $f(R)$-gravity global monopole \cite{pp6,pp7}, semiclassical gravitational effects around global monopole in Brans-Dicke theory \cite{pp8}, wormhole solutions with global monopole charge in the context of $f(Q)$-gravity \cite{pp9}, wormhole in the Milky Way galaxy with global monopole charge \cite{pp10}, conical wormholes with a global monopole charge \cite{pp11}, topologically charged four-dimensional wormholes \cite{aa50,aa53,aa54,aa55}, and higher-dimensional topologically charged wormhole \cite{MPLA}.

To prevent the formation of event horizons, the red-shift function $\Phi(r)$ must remain finite everywhere. In the specific case relevant to our wormhole solutions, we are particularly interested in scenarios where the red-shift function is constant, {\it i. e.}, $\Phi'(r)=0$. Without a loss of generality, we can consider $\Phi(r)=0$. Hence, the line-element (\ref{b1}) can be rewritten as 
\begin{equation}
    ds^2=-dt^2+\frac{dr^2}{\alpha^2\,\left(1-\frac{A(r)}{r}\right)}+r^2\,(d\theta^2+\sin^2\theta\,d\phi^2).\label{b2}   
\end{equation}

This specific space-time (\ref{b2}) is our background model which simplifies the Einstein field equations significantly and provides particularly intriguing solutions. Throughout the analysis, we consider this particular wormhole space-time (\ref{b2}) and analyze the energy conditions for different shape function forms $A(r)$ well-known in the literature. Expressing space-time (\ref{b2}) in the form $ds^2=g_{\mu\nu}\,dx^{\mu}\,dx^{\nu}$, the metric tensor $g_{\mu\nu}$ and its contravariant form are given by
\begin{eqnarray}
    g_{\mu\nu}&=&\mbox{diag}\left(-1,\,\,\frac{r}{\alpha^2\,(r-A(r))},\,\,r^2,\,\,r^2\,\sin^2 \theta\right),\nonumber\\
    g^{\mu\nu}&=&\mbox{diag}\left(-1,\,\,\alpha^2\,\left(1-\frac{A(r)}{r}\right),\,\,\frac{1}{r^2},\,\,\frac{1}{r^2\,\sin^2 \theta}\right).\label{b3}
\end{eqnarray}

The nonzero components of the Einstein’s tensor $G_{\mu\nu}$ for the metric (\ref{b2}) are given by
\begin{eqnarray}
    &&G^{t}_{t}=-\frac{(1-\alpha^2+\alpha^2\,A')}{r^2},\nonumber\\
    &&G^{r}_{r}=\frac{\alpha^2\,\left(1-\frac{A}{r}\right)-1}{r^2},\nonumber\\
    &&G^{\theta}_{\theta}=\frac{\alpha^2\,(A-r\,A')}{2\,r^3}=G^{\phi}_{\phi},\label{b4}
\end{eqnarray}
where $(')$ denotes ordinary derivative w. r. t. argument $r$. The Ricci scalar $R=g^{\mu\nu}\,R_{\mu\nu}$ is given by
\begin{equation}
    R=\frac{2\,(1-\alpha^2+\alpha^2\,A')}{r^2}.\label{b5} 
    \end{equation}

It is noted that in the limit $\alpha=1$, that is, in the absence of any global monopole charge, these non-zero components (\ref{b4}) reduces to:
\begin{eqnarray}
    G^{t}_{t}=-\frac{A'}{r^2},\qquad
    G^{r}_{r}=-\frac{A}{r^3},\qquad
    G^{\theta}_{\theta}=\frac{(A/r-A')}{2\,r^2}=G^{\phi}_{\phi}.\label{bb4}
\end{eqnarray}
Numerous authors have studied this type of wormhole model without the redshift function by selecting various forms of the shape function $A(r)$ and analyzing the resulting outcomes within the framework of general relativity.

Since the space-time (\ref{b2}) is a non-vacuum solution of the field equations with the non-zero components of the Einstein tensor given in Eq. (\ref{b4}), we consider anisotropic fluid as matter content whose energy-momentum tensor is given by the following form: 
\begin{equation}
    T^{\mu\nu}=(\rho+p_t)\,U^{\mu}\,U^{\nu}+(p_r-p_t)\, \eta^{\mu}\,\eta^{\nu}+p_t\,g^{\mu\nu},\label{b6}
\end{equation}
where $\rho$ is the energy density, $p_r$ is the radial pressure and $p_t$ is the tangential pressure, respectively. Here, $U^{\mu}$ is the velocity four-vector which takes the values $U^{\mu}=(1,0,0,0)$ such that $U^{\mu}\,U_{\mu}=-1$ and $\eta^{\mu}$ is the radial vector which takes the form $\eta^{\mu}=\left(0,\alpha\,\left(1-\frac{A}{r}\right)^{1/2},0,0\right)$ such that $\eta^{\mu}\,\eta_{\mu}=1$. Also, these vector field satisfies the relation $U^{\mu}\,\eta_{\mu}=0$.

Solving the Einstein Field Equation (\ref{a1}) using equations (\ref{b4}) and (\ref{b6}), we obtain the following physical quantities
\begin{eqnarray}
    &&\rho=\frac{(1-\alpha^2+\alpha^2\,A')}{r^2},\nonumber\\
    &&p_r=\frac{\alpha^2\,\left(1-\frac{A}{r}\right)-1}{r^2},\nonumber\\
    &&p_t=\frac{\alpha^2\,(A-r\,A')}{2\,r^3}.\label{b7}
\end{eqnarray}

Equation (\ref{b7}) provides the expression for the energy density, the radial pressure and the tangential pressure respectively of the fluid. We see that in the limit $\alpha=1$, one will find the results similar to those obtained in Morris-Thorne wormhole models \cite{aa8,aa9,aa10}. However, in this analysis, we are mainly interested on $\alpha \neq 1$ and show how this parameter $\alpha$ control the energy conditions as well as the anisotropy parameter $\Delta$ for different types of the shape functions known in the literature.

The proper radial distance for the wormhole space-time (\ref{b1}) with global monopole charge is given by
\begin{equation}
    \tilde{\ell}(r)=\pm\,\int^{r}_{r_0}\,\frac{dr}{\alpha\,\sqrt{1-A(r)/r}}=\frac{1}{\alpha}\,\ell_{MT} (r).
\end{equation}
This proper distance is equal to $\frac{1}{\alpha}$ times the proper radial distance obtained in MT-wormhole. Since the global monopole parameter lies in the ranges $0 < \alpha <1$, thus, we have, $\tilde{\ell}(r)>\ell_{MT} (r)$.

The geometrical properties of wormholes are dependent on the shape function. In literature, various shape functions $A(r)$ are defined and wormhole structures are analyzed. Below, we consider a few known shape functions $A(r)$ and examine wormhole models and to check for the validation of the energy conditions for each wormhole model in the presence of global monopole charge.

\subsection{Wormhole Model-I: $A(r)=r_0$}\label{subsec:1}

For the first case, we examine the wormhole model with global monopole charge using the following shape function \cite{FSN,aa60}
\begin{equation}
    A(r)=r_0.\label{b8}
\end{equation}

Therefore, using this shape function (\ref{b8}), the energy density, the radial pressure and the tangential pressure, respectively from Eq. (\ref{b7}) takes the following form
\begin{equation}
    \rho=\frac{1-\alpha^2}{r^2} >0,\quad\quad\quad\quad p_r=\frac{\alpha^2\,(r-r_0)}{r^3}-\frac{1}{r^2},\quad\quad\quad\quad p_t=\frac{\alpha^2\,r_0}{2\,r^3}>0.\label{b9}                    
\end{equation}

\begin{figure}[ht!]
    \includegraphics[width=0.45\linewidth]{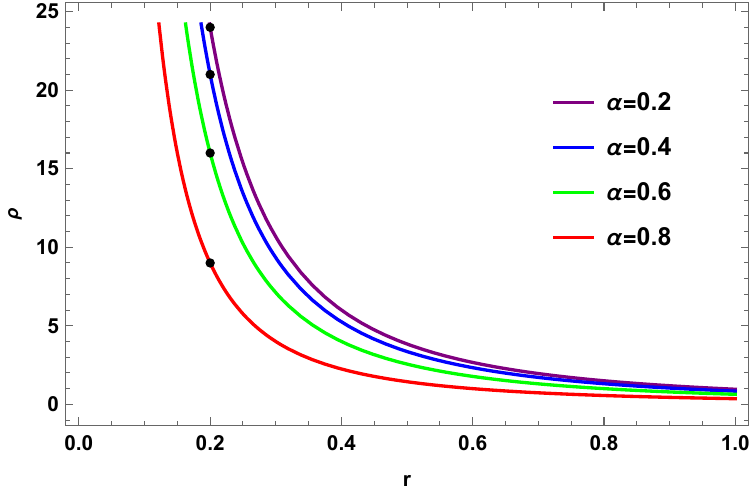}\quad\quad\quad
    \includegraphics[width=0.45\linewidth]{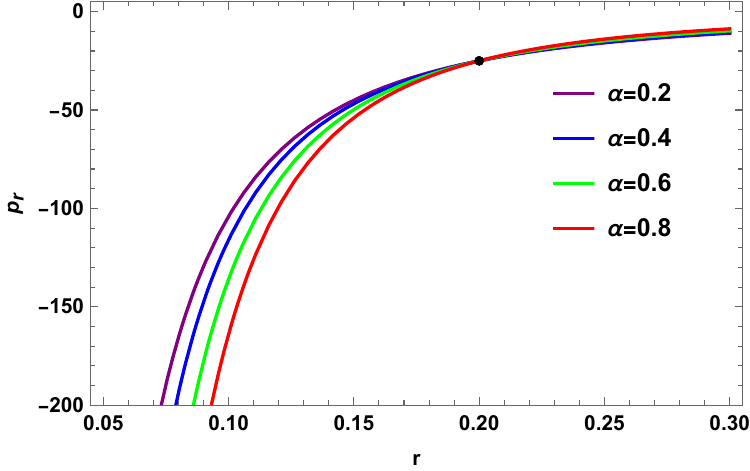}\\
    \begin{centering}
    \includegraphics[width=0.45\linewidth]{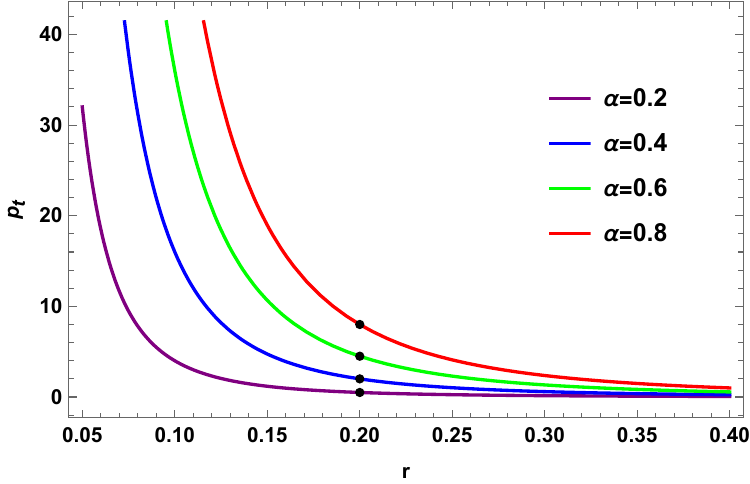}
    \par\end{centering}
    \caption{The physical quantities $\rho$ (top left), $p_r$ (top right), $p_t$ (bottom) as a function $r$ for the shape function $A(r)=r_0$ and throat radius, $r_0=0.2$.}
    \label{fig:1}
\end{figure}

We have generated graph (Figure \ref{fig:1}) of the energy density $\rho$, the radial pressure $p_r$, and the tangential pressure $p_t$ with the radial distance $r$. For that, we have fixed radius of the wormhole throat $r_0=0.2$ and choosing different values of the global monopole parameter $\alpha<1$.

\begin{figure}[ht!]
    \centering
    \includegraphics[width=0.45\linewidth]{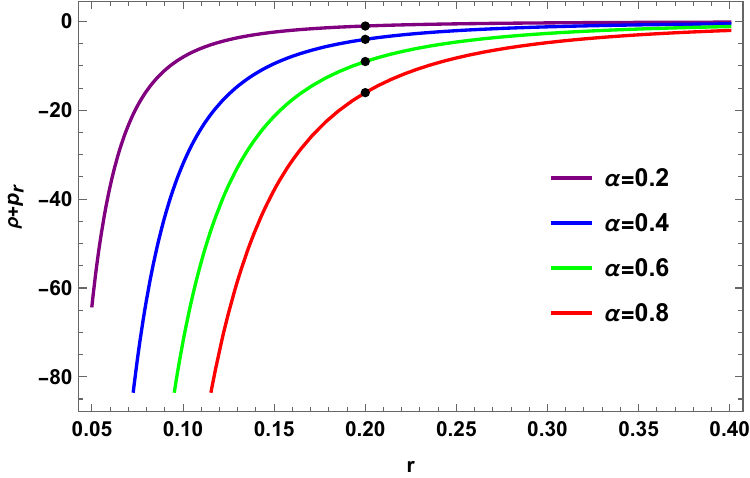}\quad\quad\quad
    \includegraphics[width=0.45\linewidth]{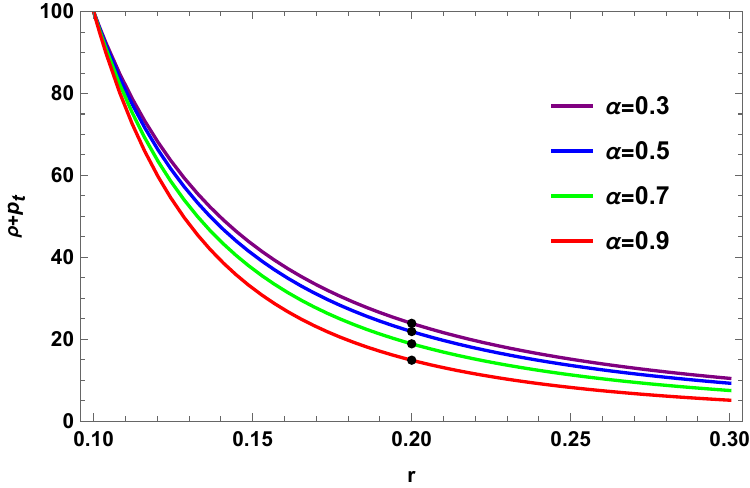}
    \caption{The behaviour of WEC terms $\rho+p_r$ (left one) and $\rho+p_t$ (right one) with $r$ for $A(r)=r_0$ and $r_0=0.2$.}
    \label{fig:2}
\end{figure}

\begin{figure}[ht!]
    \centering
    \includegraphics[width=0.45\linewidth]{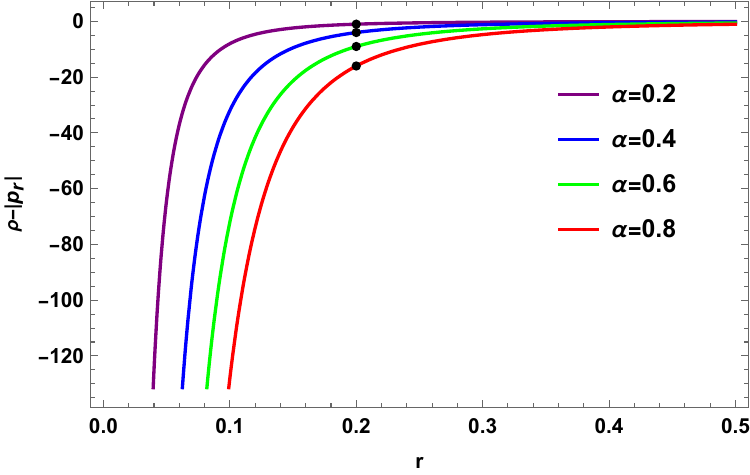}\quad\quad\quad
    \includegraphics[width=0.45\linewidth]{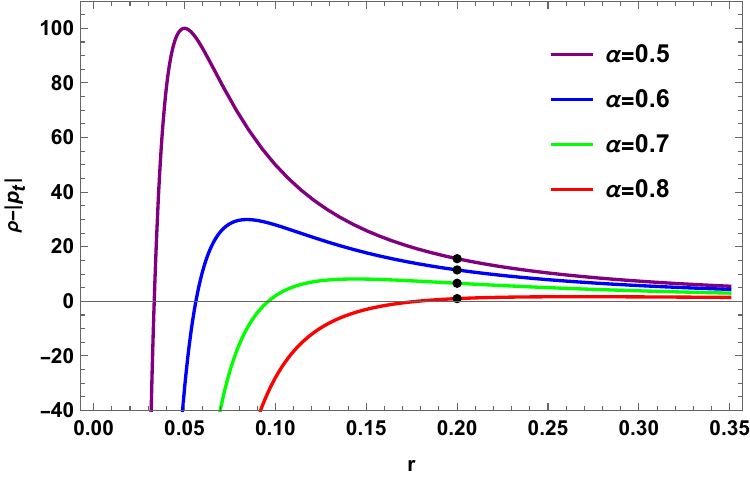}
    \caption{The behaviour of DEC terms $\rho-|p_r|$ (left one) and $\rho-|p_t|$ (right one) versus $r$ for the shape function $A(r)=r_0$ and $r_0=0.2$.}
    \label{fig:3}
\end{figure}

\begin{figure}[ht!]
    \centering
    \includegraphics[width=0.45\linewidth]{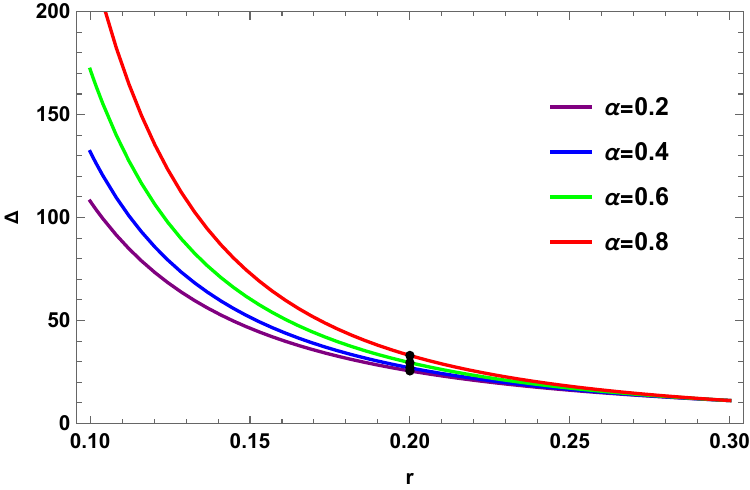}
    \caption{Anisotropy parameter ($\Delta$) versus $r$ for the shape function $A(r)=r_0$ and $r_0=0.2$.}
    \label{fig:4}
\end{figure}

From expression (\ref{b9}) of the physical quantities and Figure \ref{fig:1}, we see that the energy-density ($\rho > 0$) and the tangential pressure ($p_t>0$) are positive. While the radial pressure is negative, $p_r<0$, at the wormhole throat radius $r=r_0$. The positive energy density is control by the global monopole parameter whose value lies in the interval $0 < \alpha <1$. At the wormhole throat $r=r_0$, the physical quantities have nature $\rho(r_0)>0,\quad p_{r}(r_0)<0,\quad p_{t}(r_0)>0$ that can also be seen in Figure \ref{fig:1}. The black dot in this Figure represents the values of the corresponding physical quantities at the wormhole throat, $r_0$. From now on, in all Figures \ref{fig:2}--\ref{fig:27} throughout the analysis, the black dot on the curves will represent the values of the corresponding quantities at the wormhole throat radius, $r_0$.

Now, we discuss energy condition in this wormhole model. The first and second weak energy condition term implies 
\begin{equation}
    \text{WEC}_r:\quad \rho+p_r=-\frac{\alpha^2\,r_0}{r^3}<0,\quad\quad\quad \text{WEC}_t:\quad \rho+p_t=\left(\frac{1-\alpha^2}{r^2}+\frac{r_0\,\alpha^2}{2\,r^3}\right)>0.\label{b10}
\end{equation}

We have generated Figure \ref{fig:2} showing the behavior of $\rho+p_r$ and $\rho+p_t$ versus the radial distance $r$ keeping fixed the throat radius $r_0=0.2$. From this figure \ref{fig:2}, we see that the first weak energy condition term (as well as the NEC) is negative, $\rho+p_r<0$, in the radial direction, and the second term is positive, $\rho+p_t>0$, along the tangential direction. At the wormhole throat $r=r_0$, we have $(\rho+p_t)|_{r_0}=\frac{\left(1-\alpha^2/2\right)}{r^{2}_0}>0$. Thus, for the chosen shape function (\ref{b8}), the wormhole model with global monopole charge partially satisfied the weak energy condition.

Next, the SEC states that 
\begin{equation}
    \rho+p_r+2\,p_t=0\label{b11}
\end{equation}
which indicates that the matter content anisotropic fluid satisfied the SEC. Finally, the DECs are found to be
\begin{equation}
    \text{DEC}_r:\quad \rho-|p_r|=\frac{1-\alpha^2}{r^2}-\left|\frac{\alpha^2\,(r-r_0)}{r^3}-\frac{1}{r^2}\right|,\quad\quad \text{DEC}_t:\quad \rho-|p_t|=\frac{2-\alpha^2\,\left(2+r_0/r\right)}{2\,r^2}.\label{b12}
\end{equation}

We have generated Figure \ref{fig:3} showing the behavior of $(\rho-|p_r|)$ and $(\rho-|p_t|)$ versus the radial distance $r$ keeping fixed the wormhole throat radius $r_0=0.2$. Form this Figure, we see that the first dominant energy condition term is negative, $\rho-|p_r|<0$, while the second term $(\rho-|p_t|)>0$ for $r \geq r_0$.

At the wormhole throat $r=r_0$, the second DEC term from Eq. (\ref{b12}) is given by
\begin{equation}
    (\rho-|p_t|)|_{r_0}=\frac{2-3\,\alpha^2}{2\,r^2_{0}}.\label{b13}
\end{equation}
From Eq. (\ref{b13}), we see that the term $(\rho-|p_t|) \geq 0$ in the range of the global monopole parameter $\alpha$ given by $0 < \alpha \leq \sqrt{2/3}$, and becomes negative, $(\rho-|p_t|)<0$, in the range $\sqrt{2/3} < \alpha < 1$. Thus, for the chosen shape function (\ref{b8}), the wormhole model with global monopole charge partially satisfied the DEC along the tangential direction in the interval $0 < \alpha < \sqrt{2/3}$.

Now, we calculate the anisotropy parameter in this model and is given by
\begin{equation}
    \Delta=\alpha^2\,\left(\frac{3\,r_0}{2\,r^3}-\frac{1}{r^2}\right)+\frac{1}{r^2}.\label{b14} 
\end{equation}
We have generated Figure \ref{fig:4} of this anisotropy parameter $\Delta$ against the radial distance $r$ at the throat radius $r_0=0.2$. From this figure \ref{fig:4}, we see that $\Delta>0$ at $r=r_0$. This indicates that the wormhole geometry under consideration with the chosen shape function $A(r)=r_0$ is repulsive in nature.

Thus, we conclude that the presence of non-exotic matter is a necessary condition for the existence of wormhole solution with global monopole charge in general relativity with the shape function $A(r)=r_0$. However, we have shown that the weak and dominant energy conditions partially violated along the tangential direction.

\subsection{Wormhole Model-II:\, $A(r)=r^{2}_0/r$}\label{subsec:2}

In this model, we examine the traversable wormhole with global monopole charge using the following shape function \cite{FSN,aa60}
\begin{equation}
    A(r)=\frac{r_{0}^2}{r},\label{b15}
\end{equation}
Where $r_0$ is the throat radius. Therefore, using this shape function (\ref{b15}), we finds the energy density, radial pressure and tangential pressure from (\ref{b7}) as follows:
\begin{eqnarray}
    \rho=\frac{1}{r^2}-\frac{\alpha^2\,(r_{0}^2+r^2)}{r^4},\quad\quad\quad
    p_r=\frac{\alpha^2\,(r^2-r^{2}_0)}{r^4}-\frac{1}{r^2},\quad\quad\quad
    p_t=\frac{\alpha^2\,r^{2}_0}{r^4}.\label{b16} 
\end{eqnarray}

\begin{figure}[ht!]
    \includegraphics[width=0.45\linewidth]{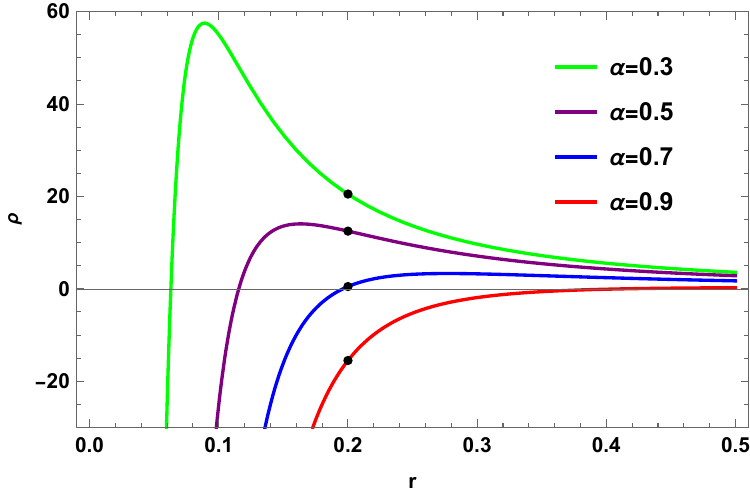}\quad\quad\quad
    \includegraphics[width=0.45\linewidth]{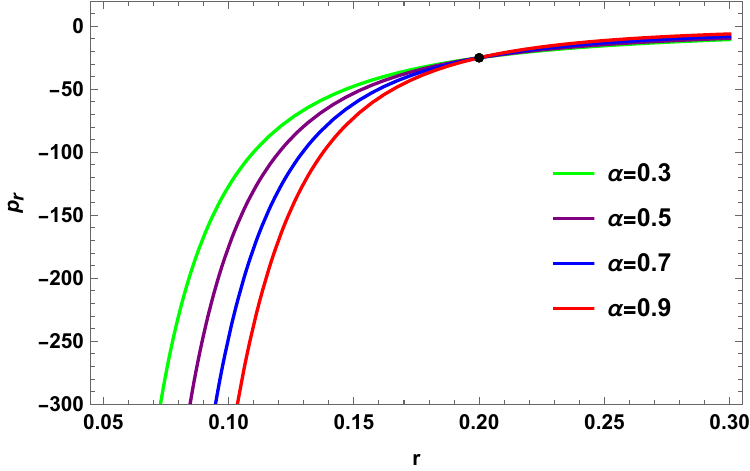}\\
    \begin{centering}
    \includegraphics[width=0.45\linewidth]{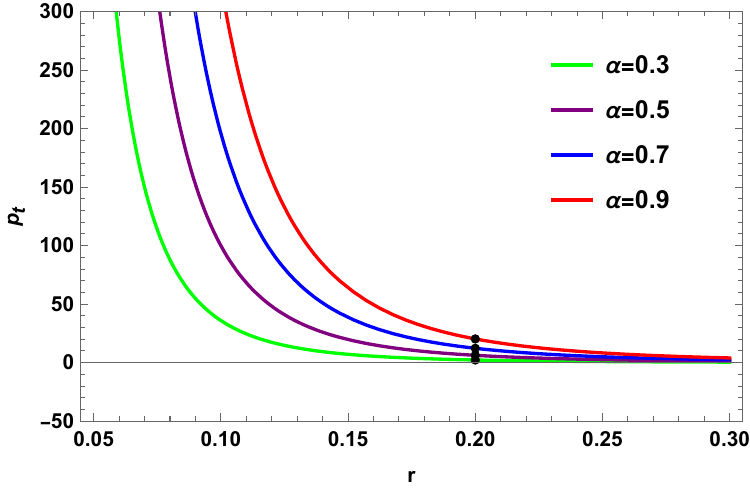}
    \par\end{centering}
    \caption{The behaviour of physical quantities $\rho$ (top left), $p_r$ (top right), $p_t$ (bottom) as a function $r$ for the shape function $A(r)=r^{2}_0/r$ and throat radius, $r_0=0.2$.}
    \label{fig:5}
\end{figure}

We have generated Figure \ref{fig:5} showing the behavior of the energy density $\rho$, the radial pressure $p_r$ and the tangential pressure $p_t$ with the radial coordinate $r$ for the throat radius $r_0=0.2$.

At the wormhole throat $r=r_0$, from Eq. (28), we finds the energy density given by
\begin{equation}
    \rho|_{r=r_0}=\frac{1-2\,\alpha^2}{r^{2}_0}.\label{b17}
\end{equation}
From the above expression (\ref{b17}), we see that the energy-density remains remains positive, $\rho>0$ in the range $0 < \alpha< 0.707$ and becomes negative, $\rho<0$ in the interval $0.707< \alpha < 1$ at the wormhole throat $r=r_0$. This point is visible in Figure \ref{fig:5}.

Now, we discuss various energy conditions for the matter content in this wormhole model. The WECs states that
\begin{eqnarray}
    \text{WEC}_r:\quad \rho+p_r=-\frac{2\,\alpha^2\,r^2_{0}}{r^4}<0,\quad\quad\quad\quad \text{WEC}_t:\quad \rho+p_t=\frac{1-\alpha^2}{r^2}>0.\label{b18}
\end{eqnarray}
From above expressions, it’s clear that the WEC is partially violated in the radial direction while satisfied in the tangential direction at the wormhole throat $r=r_0$ since the global monopole parameter lies in the range $0 < \alpha < 1$. The matter content satisfied the SEC since we finds $\rho+p_r+2\,p_t=0$.   

\begin{figure}[ht!]
    \centering
    \includegraphics[width=0.45\linewidth]{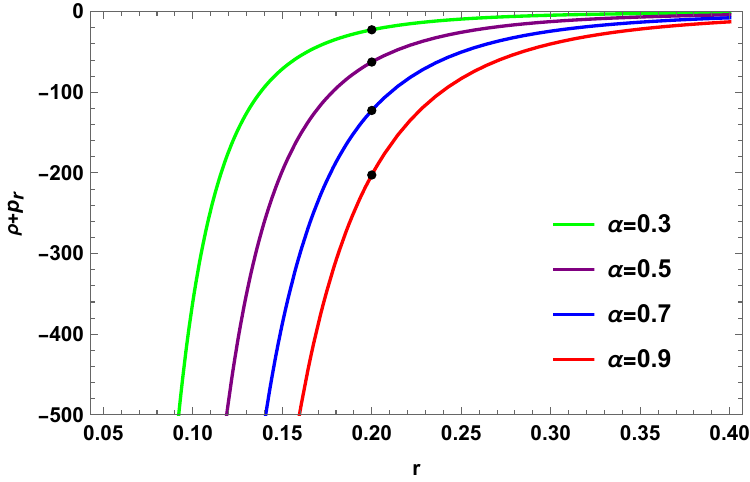}\quad\quad\quad
    \includegraphics[width=0.45\linewidth]{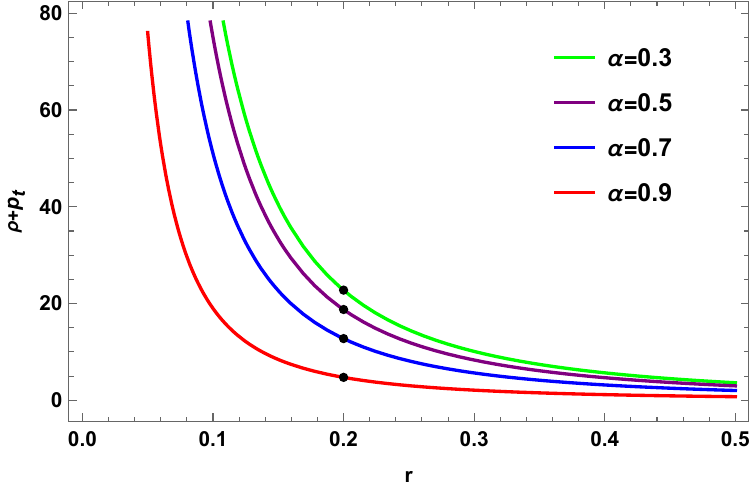}
    \caption{The behaviour of WEC terms $\rho+p_r$ (left one) and $\rho+p_t$ (right one) versus $r$ for the shape function $A(r)=r^{2}_0/r$ and $r_0=0.2$.}
    \label{fig:6}
\end{figure}

We have generated Figure \ref{fig:6} showing the behavior of $\rho+p_r$ and $\rho+p_t$ with the radial distance $r$ for the values of the throat radius $r_0=0.2$.

The DECs states that 
\begin{eqnarray}
    &&\text{DEC}_r:\quad \rho-|p_r|=\frac{1}{r^2}-\frac{\alpha^2\,(r_{0}^2+r^2)}{r^4}-\left|\frac{\alpha^2\,(r^2-r^{2}_0)}{r^4}-\frac{1}{r^2}\right|,\nonumber\\ &&\text{DEC}_t:\quad \rho-|p_t|=\frac{(1-\alpha^2)}{r^2}-\frac{2\,\alpha^2\,r^2_{0}}{r^4}.\label{b19}
\end{eqnarray}

\begin{figure}[ht!]
    \centering
    \includegraphics[width=0.45\linewidth]{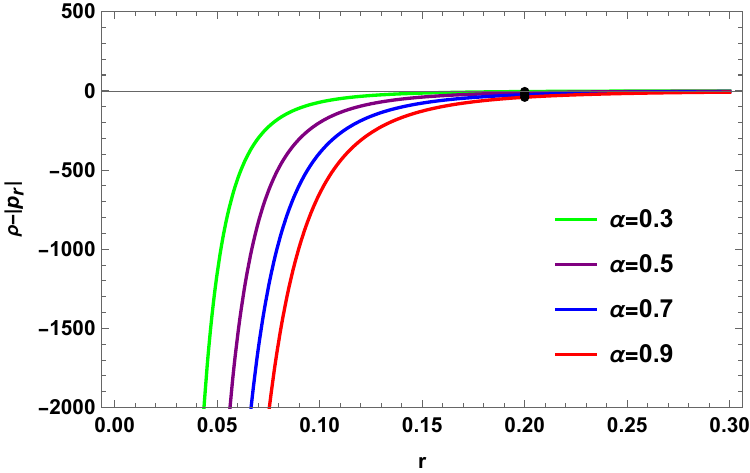}\quad\quad\quad
    \includegraphics[width=0.45\linewidth]{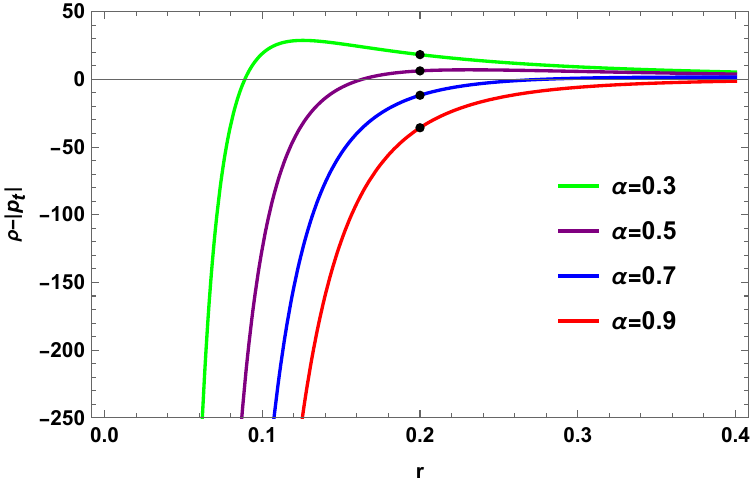}
    \caption{The nature of DEC terms $\rho-|p_r|$ (left one) and $\rho-|p_t|$ (right one) as a function of $r$ for the shape function $A(r)=r^{2}_0/r$ and $r_0=0.2$.}
    \label{fig:7}
\end{figure}

We have generated Figure \ref{fig:7} showing the behaviour of the DEC terms $\rho-|p_r|$ and $\rho-p_t|$ versus the radial distance $r$ for the throat radius $r_0=0.2$.

From Figure \ref{fig:7}, we see that the term $\rho-|p_t|$ varies between the negative and positive values for $r \geq r_0$ for different values of the global monopole parameter $\alpha<1$. At the wormhole throat, {\it i.e.,} $r=r_0$, from Eq. (\ref{b19}), we finds
\begin{equation}
    \rho-|p_t|=\frac{1-3\,\alpha^2}{r^{2}_0}.\label{b20}
\end{equation}
From Eq. (\ref{b20}), it is clear that the DEC term $\rho-|p_t|$ is positive in the range $0< \alpha <1/\sqrt{3}$ and becomes negative in $1/\sqrt{3} < \alpha < 1$. Thus, the matter content partially satisfies the DEC along the tangential direction provided the global monopole parameter lies in the range $0 < \alpha < 1/\sqrt{3}$ otherwise violates.

Finally, the anisotropy parameter in this wormhole model is given by
\begin{equation}
    \Delta=\alpha^2\,\left(\frac{2\,r^{2}_0-r^2}{r^4}\right)+\frac{1}{r^2}.\label{b21}
\end{equation}
At the wormhole throat $r=r_0$, we finds
\begin{equation}
    \Delta=\frac{1+\alpha^2}{r^{2}_0}>0.\label{b22}
\end{equation}

\begin{figure}[ht!]
    \centering
    \includegraphics[width=0.5\linewidth]{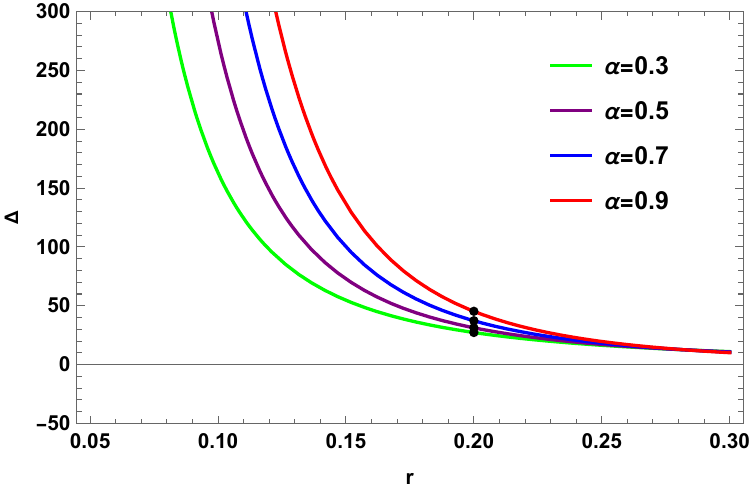}
    \caption{The nature of anisotropy parameter $\Delta$ as a function of $r$ for the chosen shape function $A(r)=r^{2}_0/r$ and $r_0=0.2$.}
    \label{fig:8}
\end{figure}

We have generated Figure \ref{fig:8} showing the behavior of this anisotropy parameter $\Delta$ as a function of the radial distance r in this model. Form this figure, we see that the wormhole geometry under consideration in model-{\bf II} is repulsive in nature.

Thus, we conclude that for the chosen shape function $A(r)=r^{2}_0/r$, the wormhole solution with global monopole charge partially satisfied the WEC and DEC provided the global monopole parameter is adjusted.

\subsection{Wormhole Model-III:\,  $A(r)=r_0\,\left[1+\gamma\,\left(1-\frac{r_0}{r}\right)\right]$}\label{subsec:3}

In this model, we examine the wormhole model using the following shape function \cite{FSN,aa60,aa65,aa63}
\begin{equation}
    A(r)=r_0\,\left[1+\gamma\,\left(1-\frac{r_0}{r}\right)\right],\label{b23}
\end{equation}
where $0 < \gamma <1$ otherwise the flare-out conditions will not satisfy.

Therefore, using this shape function (\ref{b23}), we finds the energy density, radial pressure and tangential pressure from Eq. (\ref{b7}) as follows: 
\begin{eqnarray}
    &&\rho=\frac{r^2-r^2\,\alpha^2+r^{2}_0\,\alpha^2\,\gamma}{r^4},\nonumber\\
    &&p_r=\frac{\alpha^2\,(r-r_0)(r-r_0\,\gamma)}{r^4}-\frac{1}{r^2},\nonumber\\
    &&p_t=\frac{r_0\,\alpha^2\,(r-2\,r_0\,\gamma+r\,\gamma)}{2\,r^4}.\label{b24}
\end{eqnarray}

\begin{figure}[ht!]
    \includegraphics[width=0.45\linewidth]{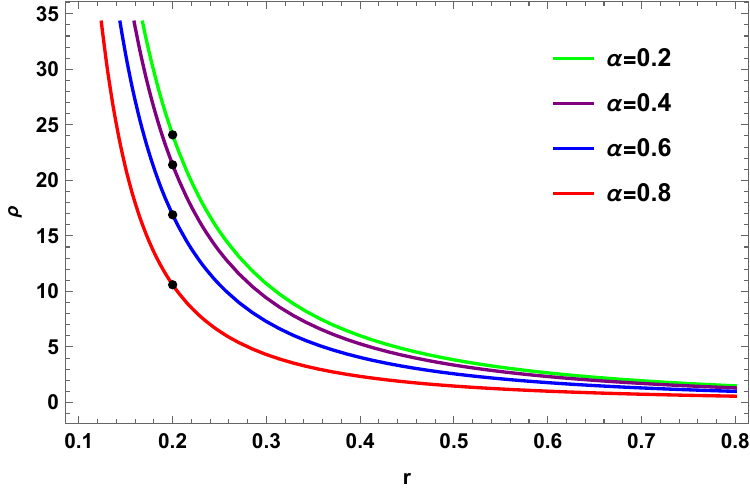}\quad\quad\quad
    \includegraphics[width=0.45\linewidth]{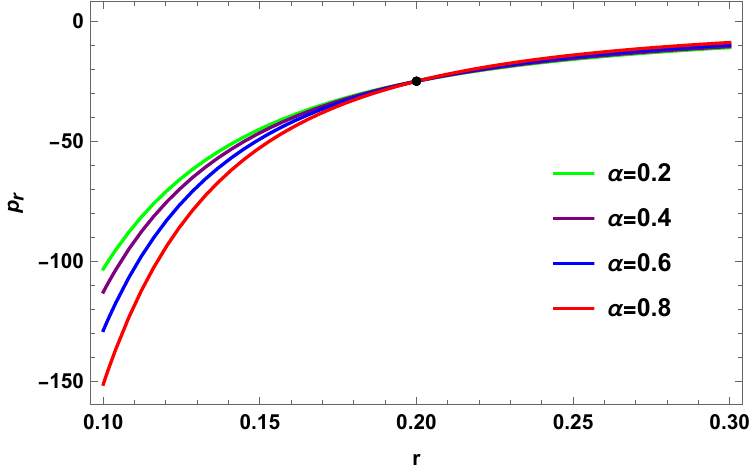}\\
    \begin{centering}
    \includegraphics[width=0.45\linewidth]{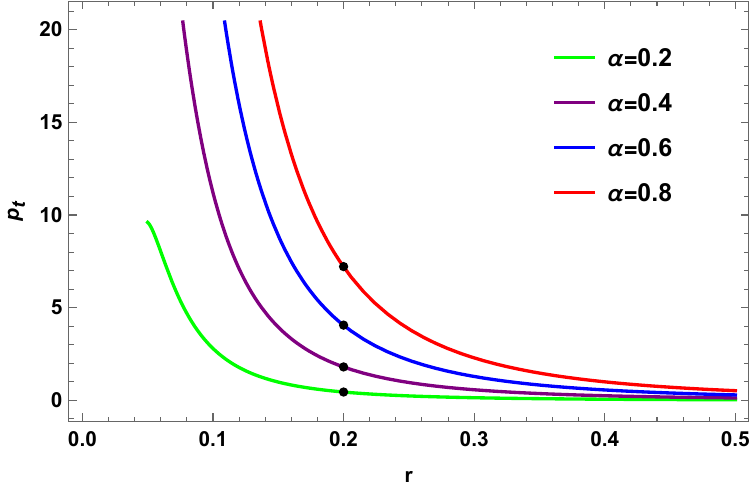}
    \par\end{centering}
    \caption{The physical quantities $\rho$ (top left), $p_r$ (top right), $p_t$ (bottom) as a function $r$ for the shape function $A(r)=r_0\,\left[1+\gamma\,\left(1-\frac{r_0}{r}\right)\right]$ with $\gamma=0.1$ and the throat radius, $r_0=0.2$.}
    \label{fig:9}
\end{figure}

We have generated Figure \ref{fig:9} showing the behavior of the energy density, $\rho,$ the radial pressure, $p_r$, and the tangential pressure, $p_t$ with the radial distance $r$ for the throat radius $r_0=0.2$ and $\gamma=0.1$.

From Figure \ref{fig:9}, we see that the energy-density ($\rho$) and the tangential pressure ($p_t$) are positive, for $r \geq r_0$. However, the radial pressure remains negative, $p_r<0$ $r \geq r_0$.

At the wormhole throat, $r=r_0$, we finds 
\begin{equation}
    \rho=\frac{1-\alpha^2\,(1-\gamma)}{r^{2}_0}>0,\quad\quad\quad p_r=-\frac{1}{r^{2}_0}<0,\quad\quad\quad p_t=\frac{\alpha^2\,(1-\gamma)}{r^{2}_0}>0.\label{b25}
\end{equation}
From the above equation, we see that the energy density ($\rho$) and the tangential pressure $(p_t)$ are positive while the radial pressure $(p_r)$ is negative at the wormhole throat $r=r_0$.

Now, we discuss various energy conditions in this model. The WECs implies that
\begin{eqnarray}
    &&\text{WEC}_r:\quad \rho+p_r=\frac{r_0\,\alpha^2\,[\gamma\,(2\,r_0-r)-r]}{r^4},\nonumber\\
    &&\text{WEC}_t:\quad \rho+p_t=\frac{2\,r^2\,(1-\alpha^2)+r\,r_0\,\alpha^2\,(1+\gamma)}{2\,r^4}.\label{b26}
\end{eqnarray}

\begin{figure}[ht!]
    \centering
    \includegraphics[width=0.45\linewidth]{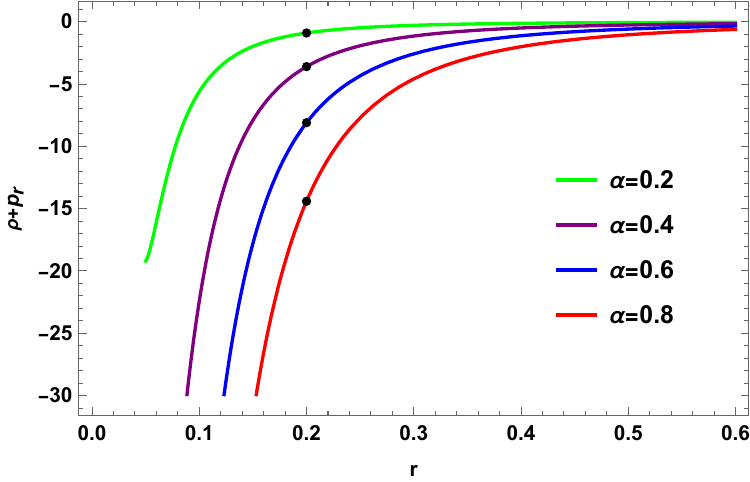}\quad\quad\quad
    \includegraphics[width=0.45\linewidth]{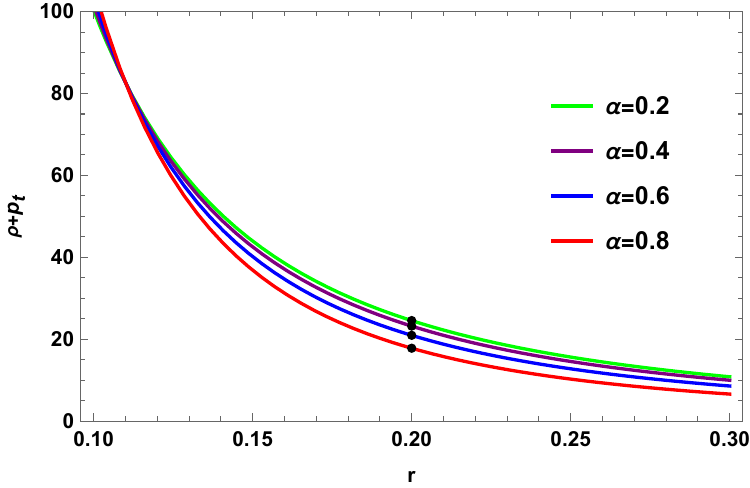}
    \caption{The nature of WEC terms $\rho+p_r$ (left one) and $\rho+p_t$ (right one) versus $r$ for the shape function $A(r)=r_0\,\left[1+\gamma\,\left(1-\frac{r_0}{r}\right)\right]$, $\gamma=0.1$ and $r_0=0.2$.}
    \label{fig:10}
\end{figure}

We have generated Figure \ref{fig:10} showing the behavior of $\rho+p_r$ and $\rho+p_t$ as a function of the radial distance $r$ for the throat radius $r_0=0.2$ and $\gamma=0.1$.

\begin{figure}[ht!]
    \centering
    \includegraphics[width=0.45\linewidth]{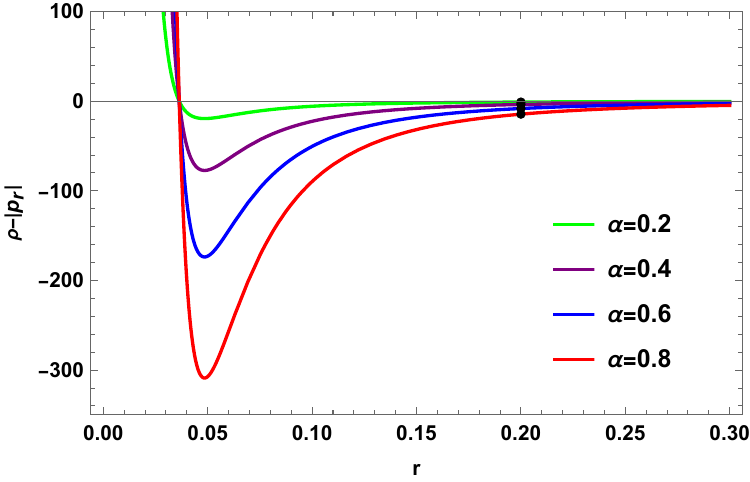}\quad\quad
    \includegraphics[width=0.45\linewidth]{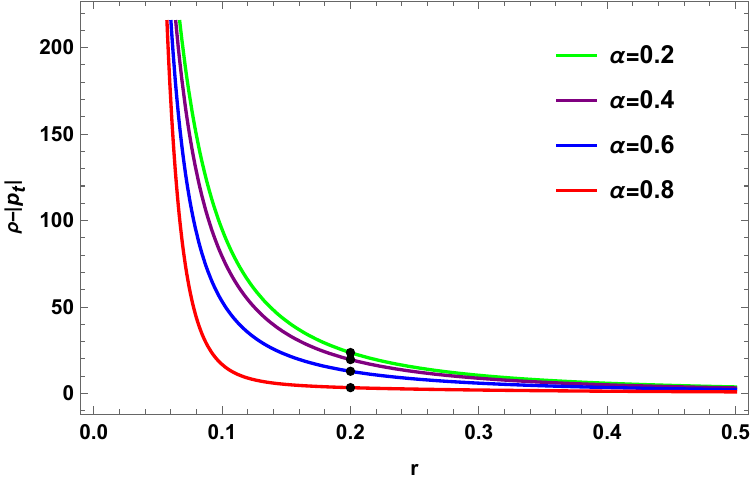}
    \caption{The nature of DEC terms $\rho-|p_r|$ (left one) and $\rho-|p_t|$ (right one) as a function $r$ for the shape function $A(r)=r_0\,\left[1+\gamma\,\left(1-\frac{r_0}{r}\right)\right]$ with $\gamma=0.1$ and the throat radius, $r_0=0.2$.}
    \label{fig:11}
\end{figure}

\begin{figure}[ht!]
    \centering
    \includegraphics[width=0.45\linewidth]{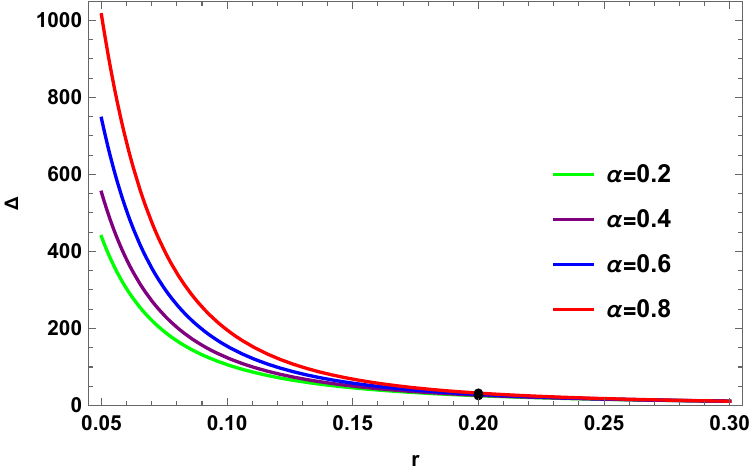}
    \caption{The behaviour of anisotropy parameter $\Delta$ as a function of $r$ for the shape function $A(r)=r_0\,\left[1+\gamma\,\left(1-\frac{r_0}{r}\right)\right]$, $\gamma=0.1$ and $r_0=0.2$. }
    \label{fig:12}
\end{figure}

From this plot \ref{fig:10}, we see that the quantity the WEC term $\rho+p_t>0$ while $\rho+p_r$ varies between positive and negative values for $r \geq r_0$.

At the wormhole throat, i.e. $r=r_0$, from Eq. (\ref{b26}) we finds
\begin{eqnarray}
    \rho+p_r=\frac{\alpha^2\,(\gamma-1)}{r^{2}_0}<0,\quad\quad\quad
    \rho+p_t=\frac{2-\alpha^2\,(1-\gamma)}{2\,r^{2}_0}>0.\label{b27}
\end{eqnarray}

From Eq. (\ref{b27}), we see that the quantity the WEC term $\rho+p_r< 0$ at the wormhole throat $r=r_0$. Thus, the WEC is satisfied in the tangential direction and violated in the radial direction. Moreover, the matter-content satisfied the SEC given by $\rho+p_r+2\,p_t=0$ throughout the space.

In addition, The DECs states that
\begin{eqnarray}
    &&\text{DEC}_r:\quad \rho-|p_r|=\frac{r^2-r^2\,\alpha^2+r^{2}_0\,\alpha^2\,\gamma}{r^4}-\left|\frac{\alpha^2\,(r-r_0)(r-r_0\,\gamma)}{r^4}-\frac{1}{r^2}\right|,\nonumber\\
    &&\text{DEC}_t:\quad \rho-|p_t|=\frac{r^2-r^2\,\alpha^2+r^{2}_0\,\alpha^2\,\gamma}{r^4}-\left|\frac{r_0\,\alpha^2\,(r-2\,r_0\,\gamma+r\,\gamma)}{2\,r^4}\right|.\label{b28}
\end{eqnarray}
At the wormhole throat, i.e., $r=r_0$, we finds from Eq. (\ref{b28}) that
\begin{eqnarray}
    \rho-|p_r|=\frac{2-\alpha^2\,(1-\gamma)}{r^2_{0}},\quad\quad\quad \rho-|p_t|=\frac{1-2\,\alpha^2\,(1-\gamma)}{r^{2}_0}.\label{b29}
\end{eqnarray}

We have generated Figure \ref{fig:11} showing the nature of the DEC terms $\rho-|p_r|$ and $\rho-|p_t|$ as a function of $r$ for the throat radius $r_0=0.2$ and $\gamma=0.1$. We see that at the throat $r_0=0.2$ and $\gamma=0.1$, the DEC term $\rho-|p_t|$ is positive and other term $\rho-|p_r|$ is negative.

Finally, the anisotropy parameter is given by
\begin{equation}
    \Delta=\frac{1-\alpha^2}{r^2}+\frac{3\,r_0\,\alpha^2\,(1+\gamma)}{2\,r^3}-\frac{2\,r^{2}_0\,\alpha^2\,\gamma}{r^4}. \label{b30}
\end{equation}
At the wormhole throat, {\it i.e.}, $r=r_0$, Eq. (\ref{b30}) becomes
\begin{equation}
    \Delta=\frac{2+\alpha^2\,(1-\gamma)}{r^{2}_0}>0.\label{b31}
\end{equation}

We have plotted $\Delta$ versus the radial distance $r$ for the throat radius $r_0=0.2$ and $\gamma=0.1$ in Figure \ref{fig:12}. From this graph, we see that $\Delta>0$, i.e., the wormhole geometry under investigation is repulsive in nature.

Thus, we conclude that for the chosen shape function $A(r)=r_0\,\left[1+\gamma\,\left(1-\frac{r_0}{r}\right)\right]$, the wormhole model with global monopole charge partially satisfied the WEC and DEC and have positive energy density.

\subsection{Wormhole Model-IV:\, $A(r)=r_0\,\frac{\cosh r_0}{\cosh r}$.}\label{subsec:4}

Here, we examine the wormhole space-time using the following shape function \cite{bb8,aa61}
\begin{equation}
    A(r)=r_0\,\left(\frac{\cosh r_0}{\cosh r}\right),\label{b32}
\end{equation}
Where $r_0$ is the throat radius. We see that at $r=r_0$, the shape  function becomes $A(r=r_0 )=r_0$.

Therefore, using this shape function (\ref{b32}), we finds the energy density, the radial pressure, and the tangential pressure from Eq. (\ref{b7}) as follows:
\begin{eqnarray}
    &&\rho=\frac{1-\alpha^2-r_0\,\alpha^2\,\cosh r_0\,\mbox{sech} r\,\tanh r}{r^2},\nonumber\\
    &&p_r=\frac{\alpha^2\,(r-r_0\,\cosh r_0\,\mbox{sech} r)}{r^3}-\frac{1}{r^2},\nonumber\\
    &&p_t=\frac{\alpha^2\,r_0\,\cosh r_0\,\mbox{sech} r\,\left(1/r+\tanh r\right)}{2\,r^2}.\label{b33}
\end{eqnarray}

\begin{figure}[ht!]
    \includegraphics[width=0.45\linewidth]{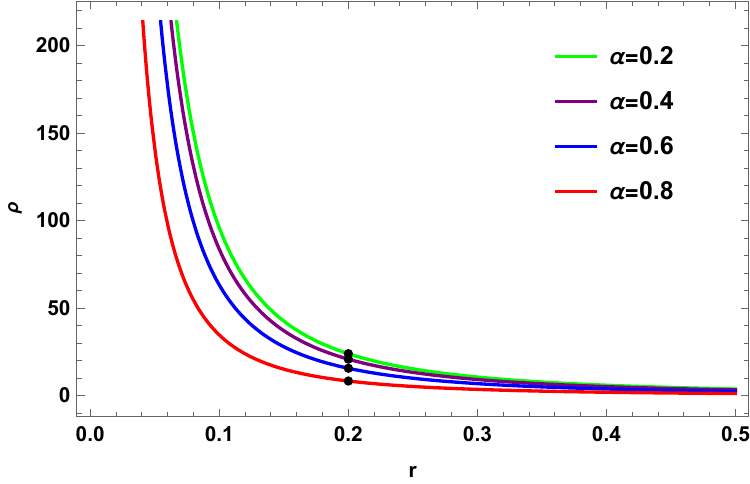}\quad\quad\quad
    \includegraphics[width=0.45\linewidth]{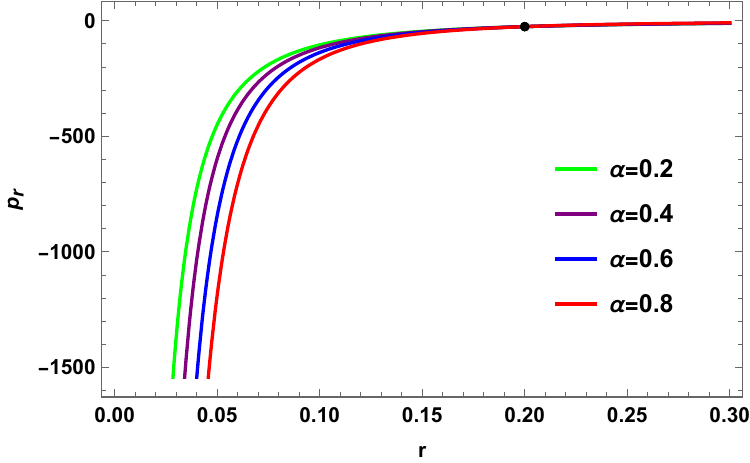}\\
    \begin{centering}
    \includegraphics[width=0.45\linewidth]{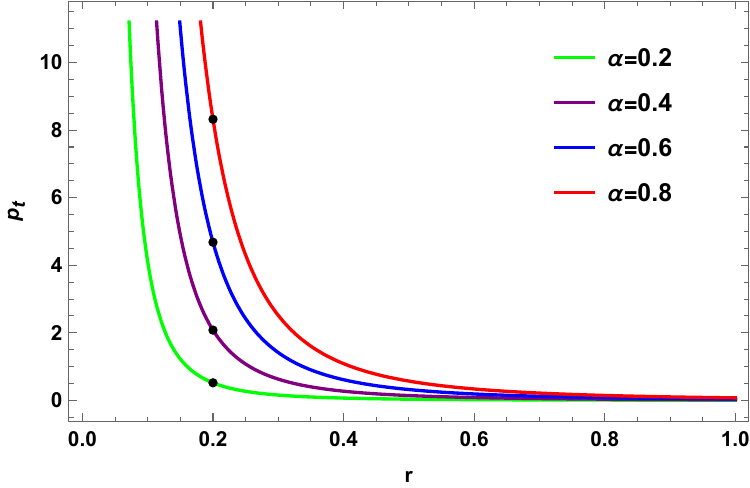}
    \par\end{centering}
    \caption{The behaviour of the physical quantities $\rho$ (top left), $p_r$ (top right), $p_t$ (bottom) as a function $r$ for the shape $A(r)=r_0\,\frac{\cosh r_0}{\cosh r}$ and the throat radius, $r_0=0.2$.}
    \label{fig:13}
\end{figure}

We have plotted the energy density, the radial pressure, and the tangential pressure of the wormhole as a function of the radial distance $r$ for the throat radius $r_0=0.2$ in Figure \ref{fig:13}.

From Figure \ref{fig:13}, we see that the energy density $\rho$ and tangential pressure $p_r$ are positive. While, the radial pressure is negative, $p_r<0$ for $r \geq r_0$.

The WECs states that
\begin{eqnarray}
    \text{WEC}_r:\quad &&\rho+p_r=-\frac{\alpha^2\,r_0\,\mbox{sech} r_0\,\mbox{sech} r\,(1+r\,\tanh r)}{2\,r^3}<0,\nonumber\\
    \text{WEC}_t:\quad &&\rho+p_t=\frac{1-\alpha^2}{r^2}+\frac{\alpha^2\,r_0\,\cosh r_0\,\mbox{sech} r\,(1-r\,\tanh r)}{2\,r^3}.\label{b34}
\end{eqnarray}

We have generated the following Figure \ref{fig:14} for the WEC terms $\rho+p_r$ and $\rho+p_t$ versus the radial distance $r$ for the throat radius $r_0=0.2$.

\begin{figure}[ht!]
    \centering
    \includegraphics[width=0.42\linewidth]{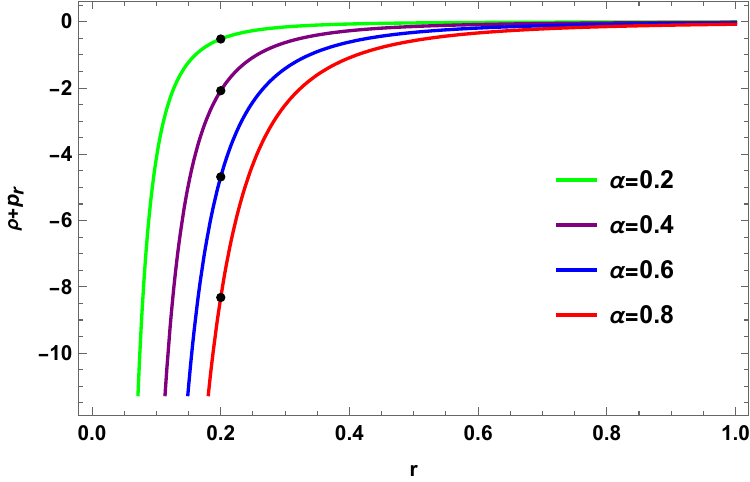}\quad\quad
    \includegraphics[width=0.42\linewidth]{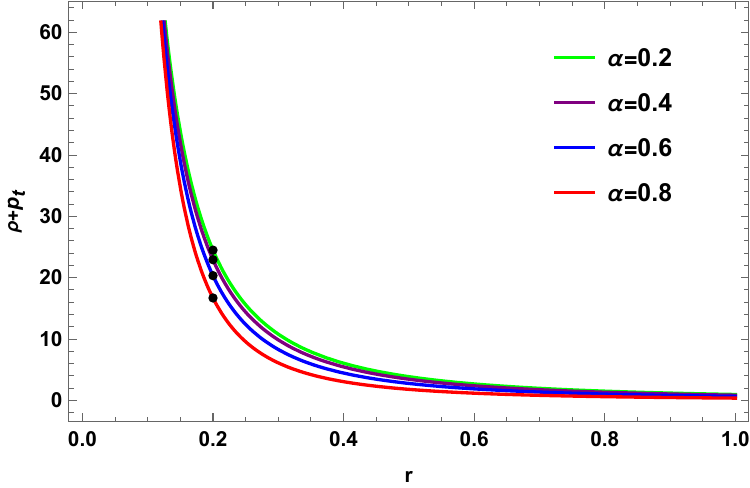}
    \caption{The behaviour of WEC terms $\rho+p_r$ (left one) and $\rho+p_t$ (right one) as a function of $r$ for the shape function $A(r)=r_0\,\frac{\cosh r_0}{\cosh r}$ and the throat radius, $r_0=0.2$.}
    \label{fig:14}
    \hfill\\
    \centering
    \includegraphics[width=0.42\linewidth]{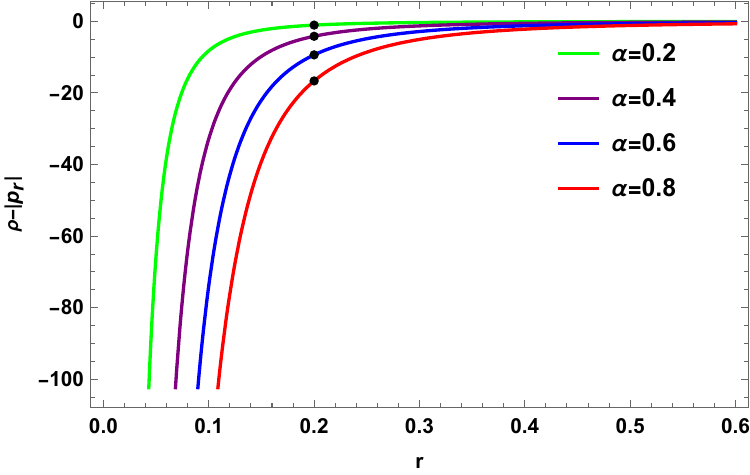}\quad\quad
    \includegraphics[width=0.42\linewidth]{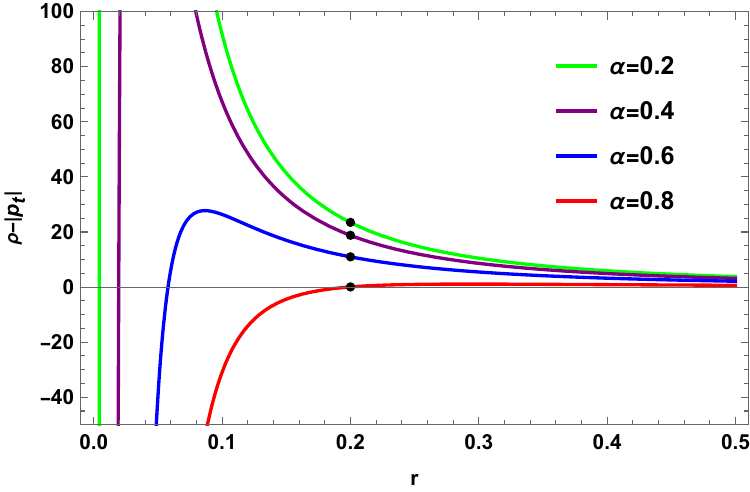}
    \caption{The behaviour of DEC terms $\rho-|p_r|$ (left one) and $\rho-|p_t|$ (right one) as a function of $r$ for the shape function $A(r)=r_0\,\frac{\cosh r_0}{\cosh r}$ and the throat radius, $r_0=0.2$.}
    \label{fig:15}
\end{figure}

From Figure \ref{fig:14}, we have observed that $\rho+p_r<0$ while $\rho+p_t>0$ for different values of the global monopole parameter $\alpha$. In absence of global monopole effect, the quantity ($\rho+p_t)<0$. Hence, the WEC is violated in the radial direction and satisfied in the tangential direction under the influence of global monopole parameter $\alpha$. 

Next, the matter content satisfies the SEC since $\rho+p_r+2\,p_t=0$. Finally, the DEC states that
\begin{eqnarray}
    \text{DEC}_r:&&\rho-|p_r|=\frac{1-\alpha^2-r_0\,\alpha^2\,\cosh r_0\,\mbox{sech} r\,\tanh r}{r^2}-\left|\frac{1-\alpha^2-r_0\,\alpha^2\,\cosh r_0\,\mbox{sech} r\,\tanh r}{r^2}\right|,\nonumber\\
    \text{DEC}_t:&&\rho-|p_t|=\frac{1-\alpha^2-r_0\,\alpha^2\,\cosh r_0\,\mbox{sech} r\,\tanh r}{r^2}-\left |\frac{\alpha^2\,r_0\,\cosh r_0\,\mbox{sech} r\,\left(1/r+\tanh r\right)}{2\,r^2}\right|.\label{b35}
\end{eqnarray}

We have generated Figure \ref{fig:15} for the quantity $\rho-|p_t|$ versus the radial distance $r$ for the throat radius $r_0=0.2$. From Figure \ref{fig:15}, we see that $\rho-|p_t|>0$ for different values of the global monopole parameter $\alpha$ except the condition where there is no effect of this parameter ($\alpha=1$). Hence, the DEC is violated in the radial direction and satisfied in the tangential direction in the presence of global monopole charge.

Finally, the anisotropy parameter is given by
\begin{equation}
    \Delta=\frac{1-\alpha^2}{r^3}+\frac{\alpha^2\,r_0\,\cosh r_0\,\mbox{sech} r\,(3+r\,\tanh r)}{2\,r^3}>0.\label{b36}
\end{equation}

\begin{figure}[ht!]
    \centering
    \includegraphics[width=0.5\linewidth]{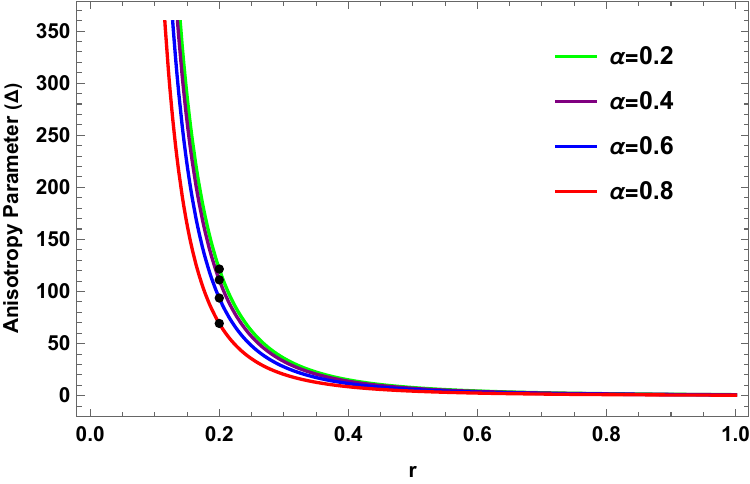}
    \caption{The behaviour of $\Delta$ versus $r$ for the shape function $A(r)=r_0\,\frac{\cosh r_0}{\cosh r}$ and the throat radius, $r_0=0.2$.}
    \label{fig:16}
\end{figure}

We have plotted this quantity $\Delta$ versus the radial distance $r$ for the throat radius $r_0=0.5$ in Figure \ref{fig:16}. From this plot, we see that $\Delta>0$ for $r \geq r_0$. Hence, the wormhole model under consideration here is repulsive in nature.

Thus, we conclude that the presence of non-exotic matter is a necessary condition for the existence of wormhole solution with global monopole charge with the shape function $A(r)=r_0\,\left(\frac{\cosh r_0}{\cosh r}\right)$. However, the matter content partially violated the weak and dominant energy conditions along the radial direction.

\subsection{Wormhole Model-V:\, $A(r)=r_0\,\frac{\tanh r}{\tanh r_0}$.}\label{subsec:5}

In this model, we examine the wormhole space-time by choosing the following shape function form given by \cite{aa64,aa62}
\begin{equation}
    A(r)=r_0\,\left(\frac{\tanh r}{\tanh r_0}\right),\label{b37}
\end{equation}
Where $r_0$ is the throat radius.

Therefore, using this shape function (\ref{b37}), we finds the energy density, the radial pressure, and the tangential pressure from Eq. (\ref{b7}) as follows:
\begin{eqnarray}
    &&\rho=\frac{1-\alpha^2}{r^2}+\frac{\alpha^2\,r_0\,\coth r_0\,\mbox{sech}^2 r}{r^2}>0,\nonumber\\
    &&p_r=\frac{\alpha^2\,(r-r_0\,\coth r_0\,\tanh r)}{r^3}-\frac{1}{r^2},\nonumber\\
    &&p_t=-\frac{\alpha^2\,r_0\,\coth r_0\,(r\,\mbox{sech}^2 r-\tanh r)}{2\,r^3}.\label{b38}
\end{eqnarray}

\begin{figure}[ht!]
    \includegraphics[width=0.45\linewidth]{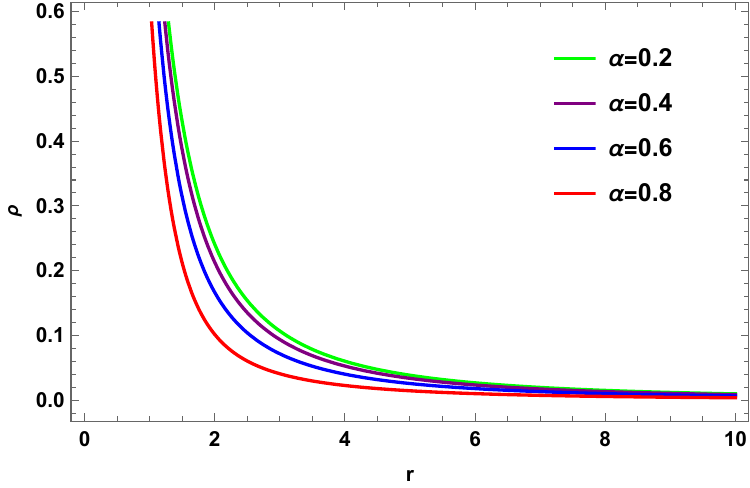}\quad\quad\quad
    \includegraphics[width=0.45\linewidth]{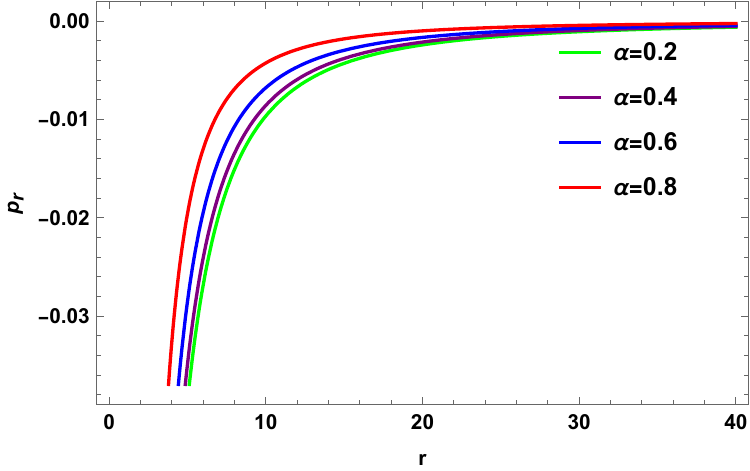}\\
    \begin{centering}
    \includegraphics[width=0.45\linewidth]{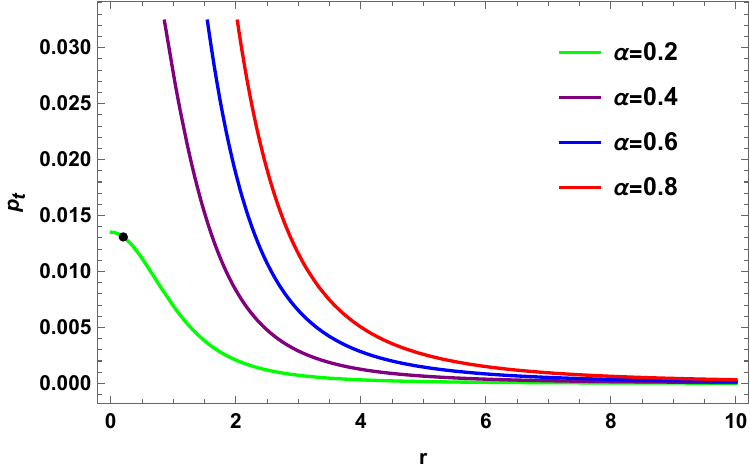}
    \par\end{centering}
    \caption{The behaviour of the physical quantities $\rho$ (top left), $p_r$ (top right), $p_t$ (bottom) as a function $r$ for the shape function $A(r)=r_0\,\frac{\tanh r}{\tanh r_0}$ and the throat radius, $r_0=0.2$.}
    \label{fig:17}
\end{figure}

We have generated Figure \ref{fig:17} for the energy density, the radial pressure and, the tangential pressure with the radial distance $r$ for the throat radius $r_0=0.2$. From this plot, we see that the energy-density ($\rho$) and the tangential pressure ($p_t$) are positive while the radial pressure is negative, $p_r<0$.

Now, we discuss various energy conditions. The WECs implies the following relations
\begin{eqnarray}
    \text{WEC}_r:\quad &&\rho+p_r=\frac{\alpha^2\,r_0\,\coth r_0\,(r\,\mbox{sech}^2 r+\tanh r)}{r^3}>0,\nonumber\\
    \text{WEC}_t:\quad &&\rho+p_t=\frac{1-\alpha^2}{r^2}+\frac{\alpha^2\,r_0\,\coth r_0\,(r\,\mbox{sech}^2 r+\tanh r)}{2\,r^3}>0.\label{b39}
\end{eqnarray}

\begin{figure}[ht!]
    \centering
    \includegraphics[width=0.45\linewidth]{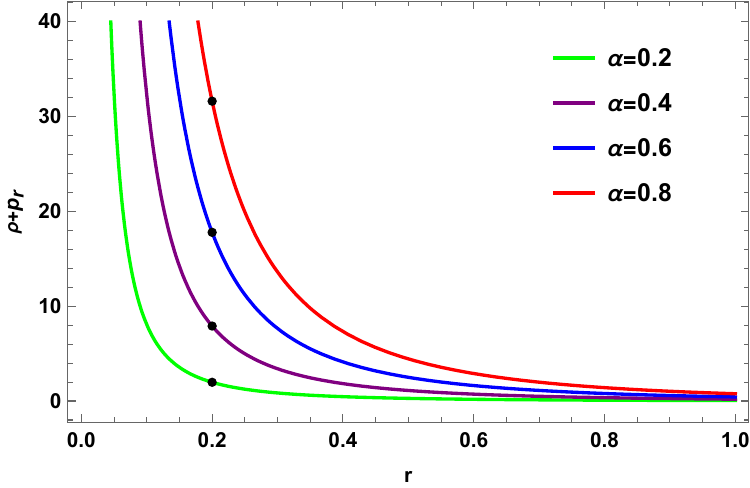}\quad\quad\quad
    \includegraphics[width=0.45\linewidth]{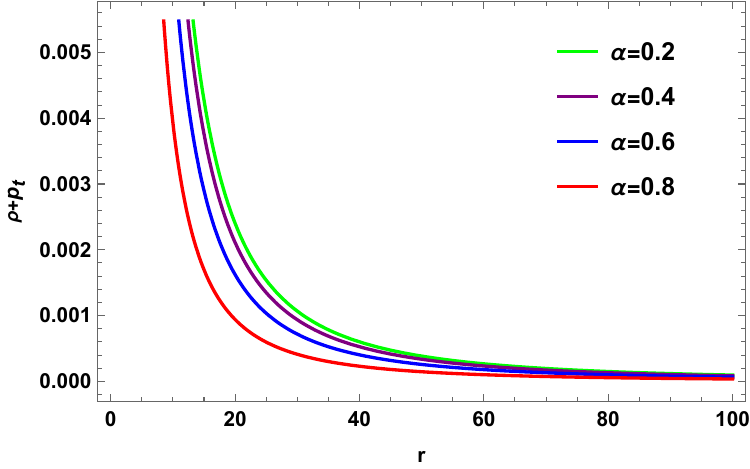}
    \caption{The behaviour of WEC terms $\rho+p_r$ (left one) and $\rho+p_t$ (right one) with $r$ for the shape function $A(r)=r_0\,\frac{\tanh r}{\tanh r_0}$ and the throat radius, $r_0=0.2$.}
    \label{fig:18}
\end{figure}

We have generated Figure \ref{fig:18} for the WEC terms quantities $\rho+p_r$ and $\rho+p_t$ with the radial distance $r$ for the throat radius $r_0=0.2$. From this graph, it’s clear that the WEC terms $\rho+p_r>0$ and $\rho+p_t>0$. Thus, we assured that the matter content anisotropic fluid fully satisfied the WEC. Moreover, the matter content satisfied the SEC because $\rho+p_r+2\,p_t=0$.

%\hfill\\
\begin{figure}[ht!]
    \centering
    \includegraphics[width=0.45\linewidth]{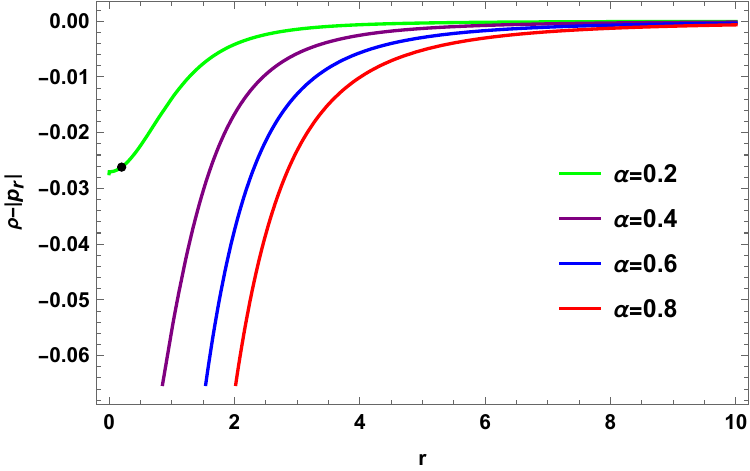}\quad\quad\quad
    \includegraphics[width=0.45\linewidth]{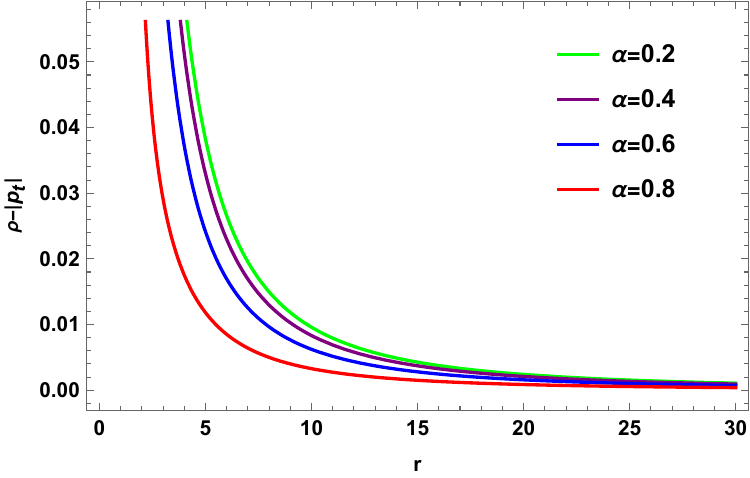}
    \caption{The behaviour of DEC terms $\rho-|p_r|$ (left one) and $\rho-|p_t|$ (right one) as a function of $r$ for the shape function $A(r)=r_0\,\frac{\tanh r}{\tanh r_0}$ and the throat radius, $r_0=0.2$.}
    \label{fig:19}
\end{figure}
%\hfill\\
\begin{figure}[ht!]
    \centering
    \includegraphics[width=0.5\linewidth]{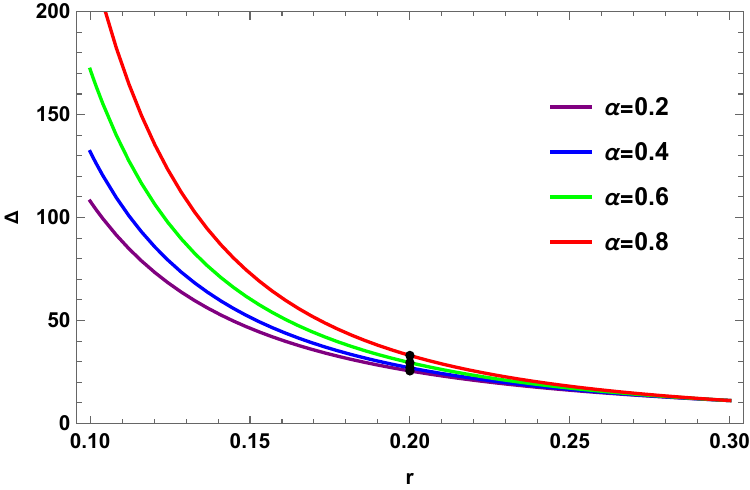}
    \caption{The behaviour of anisotropy parameter $\Delta$ with $r$ for the shape function $A(r)=r_0\,\frac{\tanh r}{\tanh r_0}$ and the throat radius, $r_0=0.2$.}
    \label{fig:20}
\end{figure}

In addition, the DECs implies the following relations
\begin{eqnarray}
    \text{DEC}_r:\quad &&\rho-|p_r|=\frac{1-\alpha^2}{r^2}+\frac{\alpha^2\,r_0\,\coth r_0\,\mbox{sech}^2 r}{r^2}-\left|\frac{\alpha^2\,(r-r_0\,\coth r_0\,\tanh r)}{r^3}-\frac{1}{r^2}\right|,\nonumber\\
    \text{DEC}_t:\quad &&\rho-|p_t|=\frac{1-\alpha^2}{r^2}+\frac{\alpha^2\,r_0\,\coth r_0\,\mbox{sech}^2 r}{r^2}-\left|-\frac{\alpha^2\,r_0\,\coth r_0\,(r\,\mbox{sech}^2 r-\tanh r)}{2\,r^3}\right|.\label{b40}
\end{eqnarray}

We have plotted $\rho-|p_t|$ with the radial distance $r$ in Figure \ref{fig:19} for the throat radius $r_0=0.2$ and different values of the global monopole parameter $\alpha$. From this Figure, we have observed that the DEC is satisfied for different values of the global monopole parameter $\alpha$.

Finally, the anisotropy parameter is given by
\begin{equation}
    \Delta=\frac{1-\alpha^2}{r^2}+\frac{\alpha^2\,r_0\,\coth r_0\,(3\,\tanh r-r\,\mbox{sech}^2 r)}{2\,r^3}.\label{b41}
\end{equation}

We have generated Figure \ref{fig:20} of the anisotropy parameter $\Delta$ with the radial distance $r$ for the throat radius $r_0=0.2$. From this Figure, we see that the anisotropy parameter $\Delta>0$ for $r \geq r_0$ and, hence, the wormhole model under consider in this case is repulsive in nature.

Thus, we conclude that for the shape function $A(r)=r_0\,\frac{\tanh r}{\tanh r_0}$, the wormhole solution with global monopole charge is an example of wormhole model in general relativity without exotic matter. 

\subsection{Wormhole Model-VI:\, $A(r)=r_0\,\tan^{-1} (k\,r)$.}\label{subsec:6}

In this model, we choose the following shape function to study topologically charged wormhole space-time \cite{aa63}
\begin{equation}
    A(r)=r_0\,\tan^{-1} (k\,r),\label{b42}
\end{equation}
where $k$ is a constant.

One can see that at $r=r_0$, the shape function should becomes $A(r=r_0)=r_0$ provided $k=\frac{\tan 1}{r_0}$. Using this shape function (\ref{b42}), we finds the energy density, the radial pressure, and the tangential pressure from Eq. (\ref{b7}) as follows:
\begin{eqnarray}
    \rho&=&\frac{1}{r^2}\,\Bigg(1-\alpha^2+\frac{\alpha^2\,r_0\,k}{1+k^2\,r^2}\Bigg),\nonumber\\
    p_r&=&\frac{\alpha^2}{r^2}\,\Bigg(1-\frac{r_0}{r}\,\tan^{-1}(k\,r) \Bigg)-\frac{1}{r^2},\nonumber\\
    p_t&=&\frac{\alpha^2}{2\,r^3}\,\Bigg(r_0\,\tan^{-1}(k\,r)-\frac{r\,r_0\,k}{1+k^2\,r^2}\Bigg).\label{b43}
\end{eqnarray}

\begin{figure}[ht!]
    \includegraphics[width=0.45\linewidth]{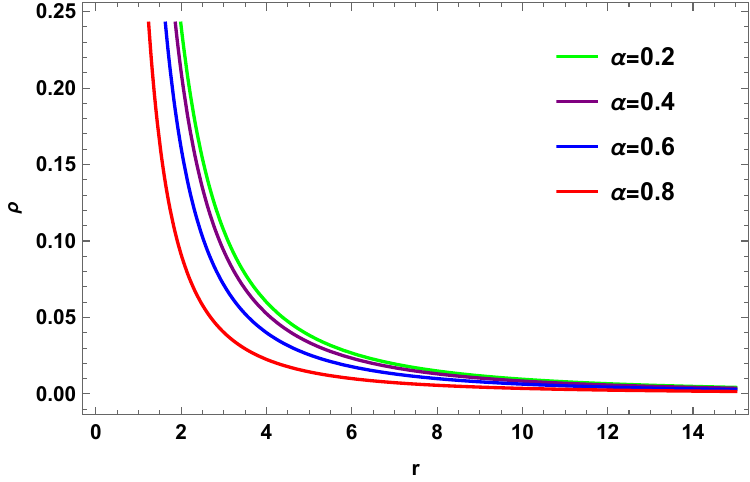}\quad\quad\quad
    \includegraphics[width=0.45\linewidth]{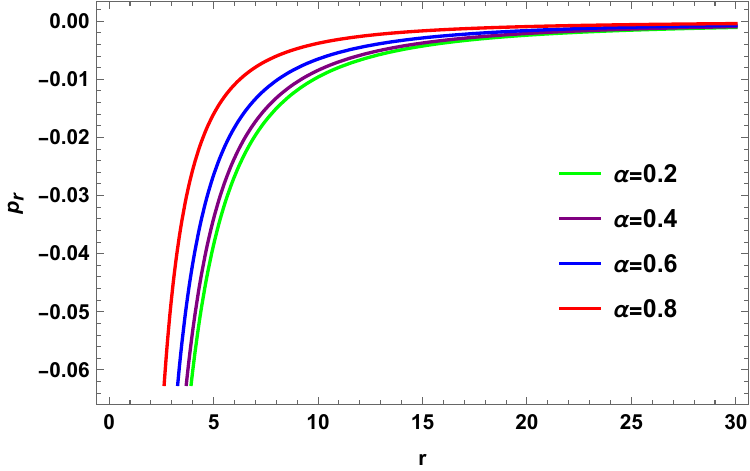}\\
    \begin{centering}
    \includegraphics[width=0.45\linewidth]{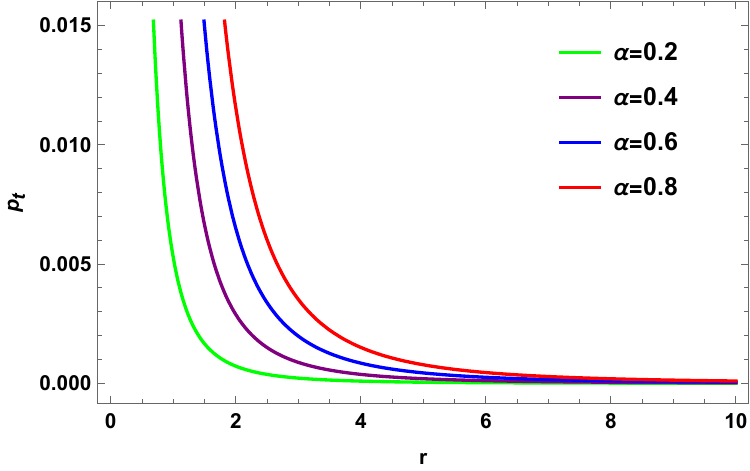}
    \par\end{centering}
    \caption{The behaviour of the physical quantities $\rho$ (top left), $p_r$ (top right), $p_t$ (bottom) as a function $r$ for the shape function $A(r)=r_0\,\tan^{-1} (k\,r)$, the throat radius, $r_0=0.2$, and the constant $k=\frac{\tan 1}{r_0}$.}
    \label{fig:21}
\end{figure}

We have generated Figure \ref{fig:21} of the energy density, the radial pressure, and the tangential pressure with the radial distance $r$ for the throat radius $r_0=0.2$ and $k=\frac{\tan 1}{r_0}$. From this graph, we see that the energy-density of matter content is positive, $\rho>0$, the radial pressure is negative, $p_r<0$, and the tangential pressure is positive, $p_t>0$ for $r \geq r_0$.

The WECs in this model implies that
\begin{eqnarray}
    \text{WEC}_r:\quad \rho+p_r&=&\frac{\alpha^2\,r_0}{r^2}\,\Bigg(\frac{k}{1+k^2\,r^2}-\frac{1}{r}\,\tan^{-1}(k\,r)\Bigg),\nonumber\\
    \text{WEC}_t:\quad \rho+p_t&=&\frac{1-\alpha^2}{r^2}+\frac{\alpha^2\,r_0}{2\,r^2}\,\Bigg(\frac{k}{1+k^2\,r^2}+\frac{1}{r}\,\tan^{-1}(k\,r)\Bigg).\label{b44}
\end{eqnarray}

\begin{figure}[ht!]
    \centering
    \includegraphics[width=0.45\linewidth]{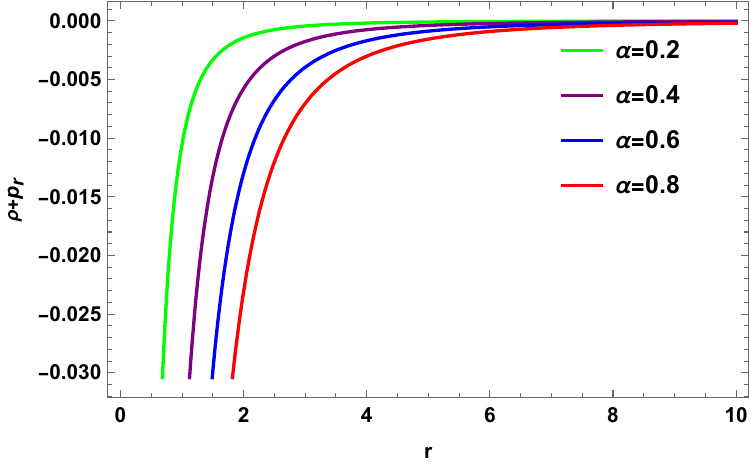}\quad\quad\quad
    \includegraphics[width=0.45\linewidth]{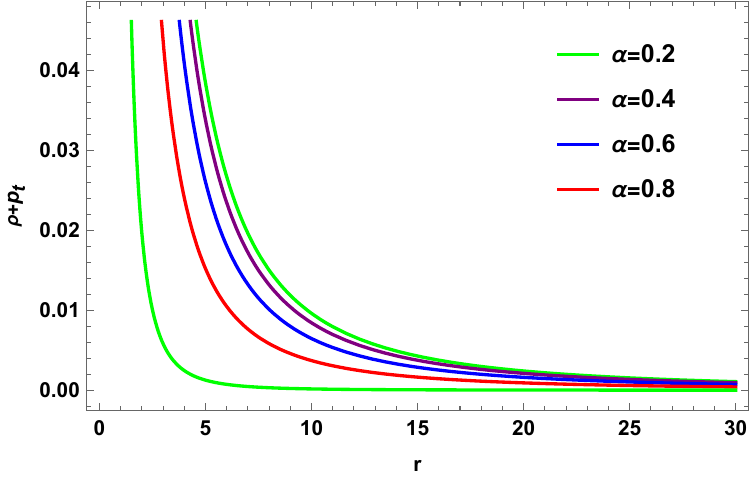}
    \caption{The behaviour of WEC terms $\rho+p_r$ (left one) and $\rho+p_t$ (right one) as a function of $r$ for the shape function $A(r)=r_0\,\tan^{-1} (k\,r)$, the throat radius, $r_0=0.2$, and the constant $k=\frac{\tan 1}{r_0}$.}
    \label{fig:22}
\end{figure}

We have generated Figure \ref{fig:22} showing the behavior of $\rho+p_r$ and $\rho+p_t$ with the radial distance $r$ for different values of the global monopole parameter keeping fixed the throat radius $r_0=0.2$ and $k=\frac{\tan 1}{r_0}$. From this figure, one can see that the WEC terms $\rho+p_r<0$ and $\rho+p_t>0$, thus, the WEC is violated in the radial direction while satisfied in the tangential direction. Moreover, our model satisfied the SEC $\rho+p_r+2\,p_t=0$.

In addition, the DECs are found to be
\begin{eqnarray}
    \text{DEC}_r: &&\rho-|p_r|=\frac{1}{r^2}\,\Bigg[1-\alpha^2+\frac{\alpha^2\,r_0\,k}{1+k^2\,r^2}\Bigg]-\left|\frac{\alpha^2}{r^2}\,\Bigg[1-\frac{r_0}{r}\,\tan^{-1}(k\,r) \Bigg]-\frac{1}{r^2}\right|,\nonumber\\
    \text{DEC}_t: &&\rho-|p_t|=\frac{1}{r^2}\,\Bigg[1-\alpha^2+\frac{\alpha^2\,r_0\,k}{1+k^2\,r^2}\Bigg]-\left|\frac{\alpha^2}{2\,r^3}\,\Bigg[r_0\,\tan^{-1}(k\,r)-\frac{r\,r_0\,k}{1+k^2\,r^2}\Bigg]\right|.\label{b45}
\end{eqnarray}

\begin{figure}[ht!]
    \centering
    \includegraphics[width=0.45\linewidth]{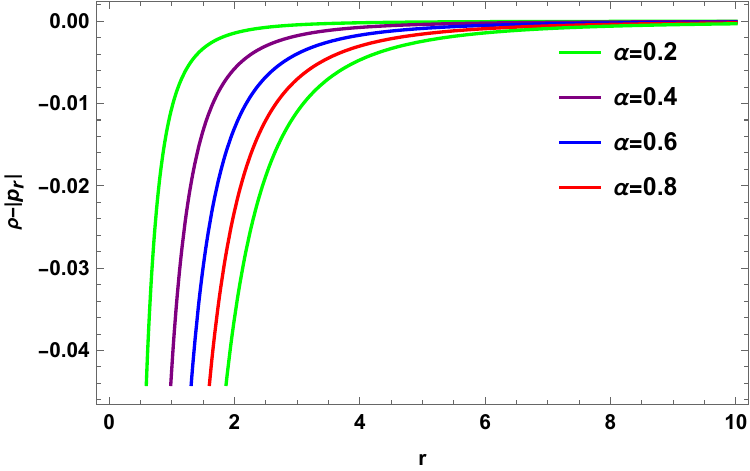}\quad\quad\quad
    \includegraphics[width=0.45\linewidth]{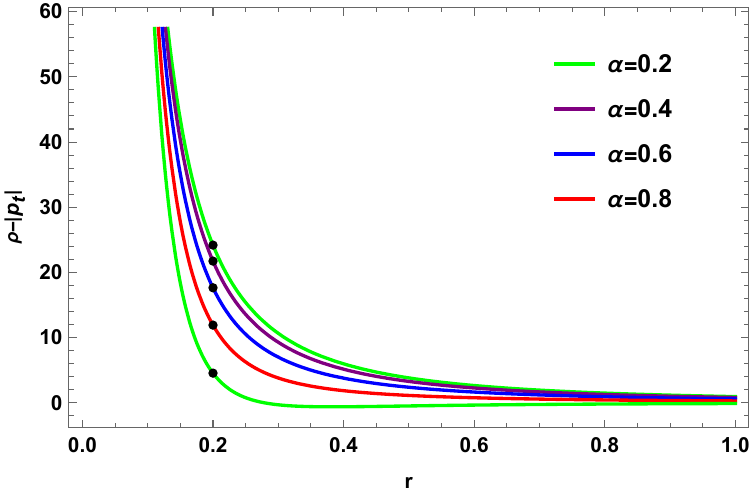}
    \caption{The behaviour of DEC terms $\rho-|p_r|$ (left one) and $\rho-|p_t|$ (right one) as a function of $r$ for the shape function $A(r)=r_0\,\tan^{-1} (k\,r)$, the throat radius, $r_0=0.2$, and the constant $k=\frac{\tan 1}{r_0}$.}
    \label{fig:23}
\end{figure}

We have plotted $\rho-p_t|$ with the radial distance $r$ in Figure \ref{fig:23} keeping the throat radius $r_0=0.2$ and $k=\frac{\tan 1}{r_0}$.

Thus, we can see that the DEC is violated in the radial direction, $\rho-|p_r|<0$ while satisfied in the tangential direction, $\rho-|p_t|>0$.

Finally, the anisotropy parameter is given by
\begin{equation}
    \Delta=\frac{1-\alpha^2}{r^2}+\frac{3}{2}\,\frac{\alpha^2\,r_0}{r^3}\,\tan^{-1} (k\,r)-\frac{\alpha^2\,r_0\,k}{1+k^2\,r^2}.\label{b46}
\end{equation}

\begin{figure}[ht!]
    \centering
    \includegraphics[width=0.5\linewidth]{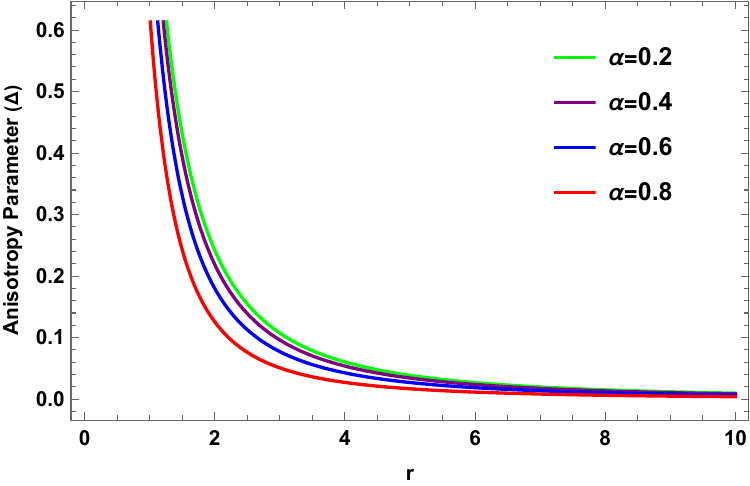}
    \caption{The behaviour of anisotropy parameter $\Delta$ with $r$ for  $A(r)=r_0\,\tan^{-1} (k\,r)$, throat radius $r_0=0.2$, and $k=\frac{\tan 1}{r_0}$.}
    \label{fig:24}
    \end{figure}

We have plotted Figure \ref{fig:24} of this parameter $\Delta$ with the radial distance $r$ keeping fixed the throat radius $r_0=0.2$, and $k=\frac{\tan 1}{r_0}$. From this Figure, one can see that this anisotropy parameter is positive, $\Delta>0$ for different values of the global monopole parameter $\alpha$, and hence, the wormhole geometry model under investigation in this case is repulsive in nature.

Thus, we conclude that the presence of non-exotic matter is a necessary condition for the existence of wormhole solution with global monopole charge in general relativity with the shape function $A(r)=r_0\,\tan^{-1} (k\,r)$. However, we have shown that the weak and dominant energy conditions are partially violated along the radial direction.

\subsection{Wormhole Model-VII:\, $A(r)=r_0\,\left(r/r_0\right)^n$.}\label{subsec:7}

In this model, we consider the following form of the shape function \cite{aa63,aa66} to examine wormholes with global monopole charge given by
\begin{equation}
    A(r)=r_0\,\left(\frac{r}{r_0}\right)^n,\label{b47}
\end{equation}
where $r_0$ is the throat radius, $n$ is an arbitrary constant with the restriction $ 0 < n< 1$ to satisfy the flaring out condition. 

Therefore, using this shape function (\ref{b47}), we finds the energy-density, the radial pressure, and the tangential pressure from Eq. (\ref{b7}) as follows:
\begin{eqnarray}
    &&\rho=\frac{1-\alpha^2}{r^2}+\frac{n\,\alpha^2}{r^2}\,\left(\frac{r}{r_0}\right)^{n-1},\nonumber\\
    &&p_r=\frac{\alpha^2}{r^2}\,\Bigg[1-\Big(\frac{r}{r_0}\Big)^{n-1}\Bigg]-\frac{1}{r^2},\nonumber\\
    &&p_t=\frac{(1-n)\,\alpha^2}{2\,r^2}\,\left(\frac{r}{r_0}\right)^{n-1}.\label{b48}
\end{eqnarray}

\renewcommand{\arraystretch}{1.5}
\begin{table}[htb!]
\centering
    \begin{tabular}{|c|c|c|c|}
    \hline
    $n$ & $\rho$ & $p_r$ & $p_t$ \\ [2.0ex] 
    \hline
    $\frac{1}{4}$ & $\frac{1-\alpha^2}{r^2}+\frac{\alpha^2}{4\,r^2}\,\left(r/r_0\right)^{-3/4}>0$ & $\frac{\alpha^2-1}{r^2}-\frac{\alpha^2}{r^2}\,\left(r/r_0\right)^{-3/4}<0$ & $\frac{3\,\alpha^2}{8\,r^2}\,\left(r/r_0\right)^{-3/4}>0$\\ [2.0ex] 
    \hline
     $\frac{1}{2}$ & $\frac{1-\alpha^2}{r^2}+\frac{\alpha^2}{2\,r^2}\,\left(r/r_0\right)^{-1/2}>0$ & $\frac{\alpha^2-1}{r^2}-\frac{\alpha^2}{r^2}\,\left(r/r_0\right)^{-1/2}<0$ & $\frac{\alpha^2}{4\,r^2}\,\left(r/r_0\right)^{-1/2}>0$ \\ [2.0ex] 
     \hline 
     $\frac{3}{4}$ & $\frac{1-\alpha^2}{r^2}+\frac{3\,\alpha^2}{4\,r^2}\,\left(r/r_0\right)^{-1/4}>0$ & $\frac{\alpha^2-1}{r^2}-\frac{\alpha^2}{r^2}\,\left(r/r_0\right)^{-1/4}<0$ & $\frac{\alpha^2}{8\,r^2}\,\left(r/r_0\right)^{-1/4}>0$ \\ [2.0ex] 
      \hline
    \end{tabular}
    \caption{The physical quantities associated with the fluid for different values of $n$. }
    \label{table:1}
\end{table}

\renewcommand{\arraystretch}{1.5}
\begin{table}[htb!]
    \centering
    \begin{tabular}{|c|c|c|}
    \hline
    $n$ & $\text{WEC}_r:\rho+p_r$ & $\text{WEC}_t:\rho+p_t$ \\ [2.0ex] 
    \hline 
    $\frac{1}{4}$ & $-\frac{3\,\alpha^2}{4\,r^2}\,\left(r/r_0\right)^{-3/4}<0$ & $\frac{1-\alpha^2}{r^2}+\frac{5\,\alpha^2}{8\,r^2}\,\left(r/r_0\right)^{-3/4}>0$ \\ [2.0ex] 
    \hline
     $\frac{1}{2}$ & $-\frac{\alpha^2}{2\,r^2}\,\left(r/r_0\right)^{-1/2}<0$ & $\frac{1-\alpha^2}{r^2}+\frac{3\,\alpha^2}{4\,r^2}\,\left(r/r_0\right)^{-1/2}>0$\\ [2.0ex] 
     \hline 
     $\frac{3}{4}$ & $-\frac{\alpha^2}{4\,r^2}\,\left(r/r_0\right)^{-1/4}<0$ & $\frac{1-\alpha^2}{r^2}+\frac{7\,\alpha^2}{8\,r^2}\,\left(r/r_0\right)^{-1/4}>0$ \\ [2.0ex] 
     \hline
    \end{tabular}
    \caption{The WEC terms for different values of $n$. }
    \label{table:2}
\end{table}

\renewcommand{\arraystretch}{1.5}
\begin{table}[htb!]
%\small
    \centering
    \begin{tabular}{|c|c|c|}
    \hline
    $n$ & $\text{DEC}_r:\rho-|p_r|$ & $\text{DEC}_t:\rho-|p_t|$ \\ [2.0ex] 
    \hline 
    $\frac{1}{4}$ & $\frac{1-\alpha^2}{r^2}+\frac{\alpha^2}{4\,r^2}\,\left(r/r_0\right)^{-3/4}-\left|\frac{\alpha^2-1}{r^2}-\frac{\alpha^2}{r^2}\,\left(r/r_0\right)^{-3/4}\right|$ & $\frac{1-\alpha^2}{r^2}-\frac{\alpha^2}{8\,r^2}\,\left(r/r_0\right)^{-3/4}$ \\ [3.0ex] 
    \hline
     $\frac{1}{2}$ & $\frac{1-\alpha^2}{r^2}+\frac{\alpha^2}{2\,r^2}\,\left(r/r_0\right)^{-1/2}-\left|\frac{\alpha^2-1}{r^2}-\frac{\alpha^2}{r^2}\,\left(r/r_0\right)^{-1/2}\right|$ & $\frac{1-\alpha^2}{r^2}+\frac{\alpha^2}{4\,r^2}\,\left(r/r_0\right)^{-1/2}$   \\ [3.0ex] 
     \hline 
     $\frac{3}{4}$ & $\frac{1-\alpha^2}{r^2}+\frac{3\,\alpha^2}{4\,r^2}\,\left(r/r_0\right)^{-1/4}-\left|\frac{\alpha^2-1}{r^2}-\frac{\alpha^2}{r^2}\,\left(r/r_0\right)^{-1/4}\right|$ & $\frac{1-\alpha^2}{r^2}+\frac{5\,\alpha^2}{8\,r^2}\,\left(r/r_0\right)^{-1/4}$ \\ [2.0ex] 
     \hline
    \end{tabular}
    \caption{The DEC terms for different values of $n$. }
    \label{table:3}
\end{table}

\begin{table}[htb!]
\centering
    \begin{tabular}{|c|c|}
    \hline
    $n$ & $\Delta=p_t-p_r$ \\ [2.0ex] 
    \hline
    $\frac{1}{4}$ & $\frac{1-\alpha^2}{r^2}+\frac{11\,\alpha^2}{8\,r^2}\,\left(r/r_0\right)^{-3/4}>0$\\ [2.0ex] 
    \hline
     $\frac{1}{2}$ & $\frac{1-\alpha^2}{r^2}+\frac{5\,\alpha^2}{4\,r^2}\,\left(r/r_0\right)^{-1/2}>0$ \\ [2.0ex] 
     \hline 
     $\frac{3}{4}$ & $\frac{1-\alpha^2}{r^2}+\frac{9\,\alpha^2}{8\,r^2}\,\left(r/r_0\right)^{-1/4}>0$\\ [2.0ex] 
      \hline
    \end{tabular}
    \caption{The anisotropy parameter for different values of $n$. }
    \label{table:4}
\end{table}

From Table \ref{table:1}, we see that the energy-density ($\rho$) of the matter content is positive for $ 0 < n < 1$. Moreover, we see from Table \ref{table:2} that the matter content partially satisfies weak energy condition along the tangential direction. Also the strong energy is automatically satisfied since $\rho+p_r+2\,p_t=0$. 

\begin{figure}[ht!]
    \includegraphics[width=0.4\linewidth]{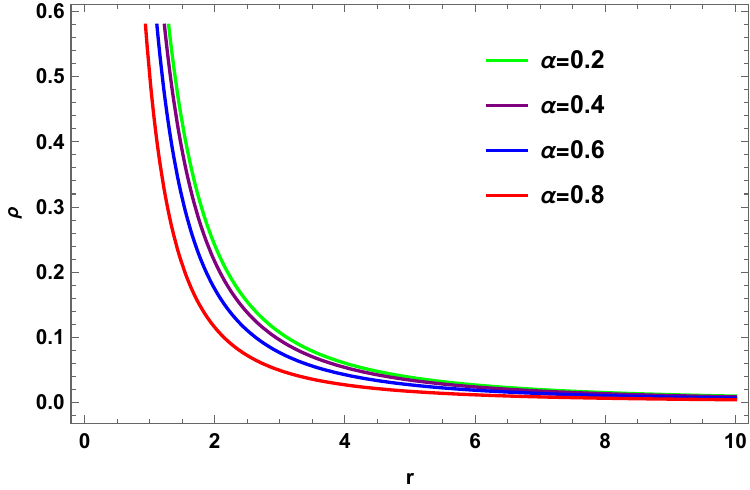}\quad\quad\quad
    \includegraphics[width=0.4\linewidth]{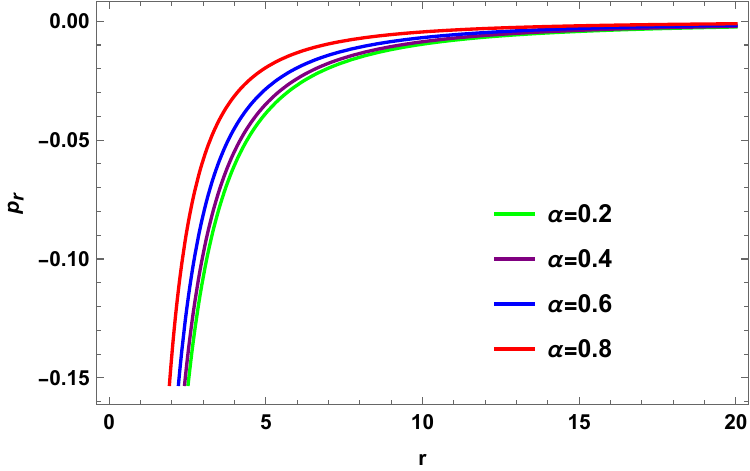}\\
    \begin{centering}
    \includegraphics[width=0.4\linewidth]{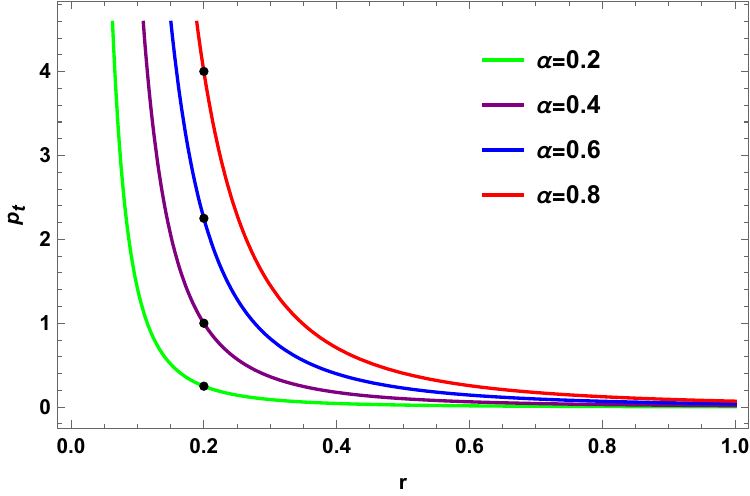}
    \par\end{centering}
    \caption{The behaviour of the physical quantities $\rho$ (top left), $p_r$ (top right), $p_t$ (bottom) as a function $r$ for the shape function $A(r)=r_0\,\left(r/r_0\right)^n$ with $n=1/2<1$, the throat radius, $r_0=0.2$.}
    \label{fig:25}
\end{figure}

\begin{figure}[ht!]
    \centering
    \includegraphics[width=0.4\linewidth]{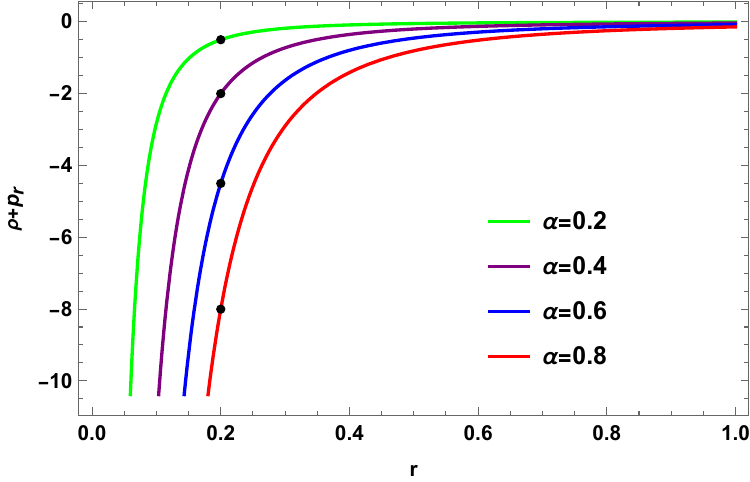}\quad\quad\quad
    \includegraphics[width=0.4\linewidth]{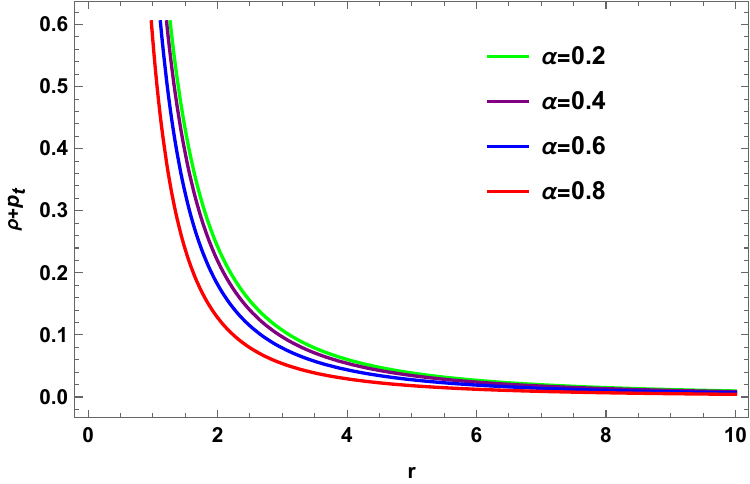}
    \caption{The behaviour of WEC terms $\rho+p_r$ (left one) and $\rho+p_t$ (right one) as a function of $r$ for the shape function $A(r)=r_0\,\left(r/r_0\right)^n$ with $n=1/2<1$, the throat radius, $r_0=0.2$.}
    \label{fig:26}
\end{figure}

We have generated Figure \ref{fig:25} for the physical quantities $\rho, p_r, p_t$ using Table \ref{table:1} for $n=1/2<1$ and Figure \ref{fig:26} for the weak energy condition using Table \ref{table:2} by choosing different values of the global monopole parameter $\alpha$.

\begin{figure}[ht!]
    \centering
    \includegraphics[width=0.4\linewidth]{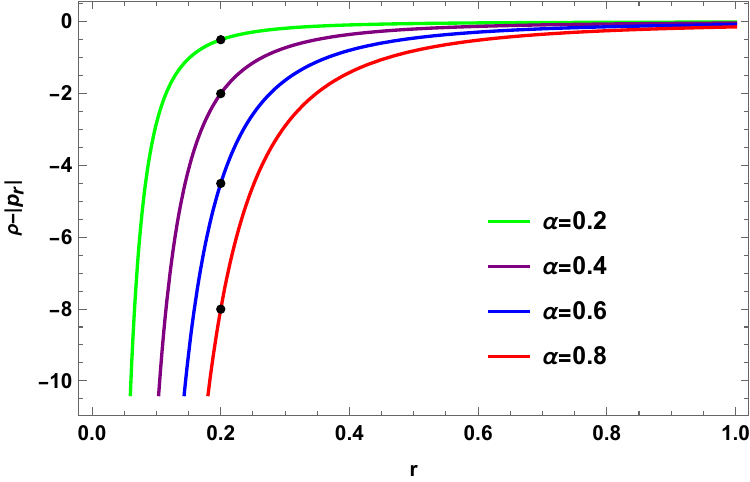}\quad\quad\quad
    \includegraphics[width=0.4\linewidth]{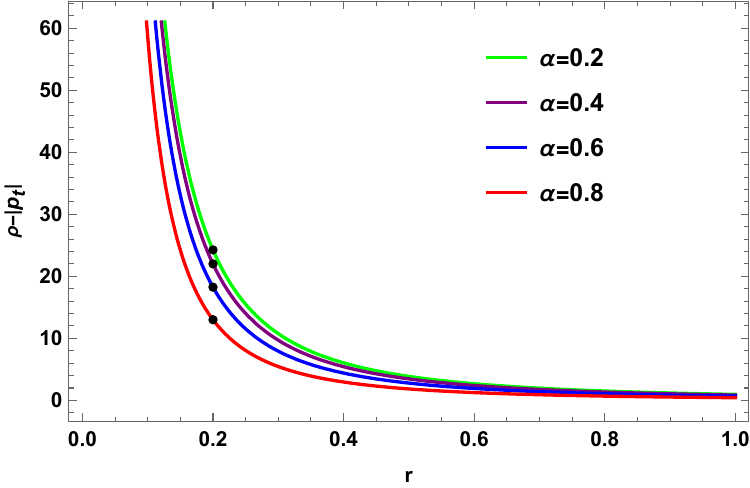}
    \caption{The behaviour of DEC terms $\rho-|p_r|$ (left one) and $\rho-|p_t|$ (right one) as a function of $r$ for the shape function $A(r)=r_0\,\left(r/r_0\right)^n$ with $n=1/2<1$, the throat radius, $r_0=0.2$.}
    \label{fig:27}
\end{figure}

Using Table \ref{table:3}, we have generated a graph of the dominant energy condition in Figure (\ref{fig:27}) for $n=1/2$ by choosing different values of global monopole parameter $\alpha$. We see that the DEC is partially satisfied along the tangential direction whereas violated in the radial direction.

%\hfil\\
%\hfill\\

\begin{figure}[ht!]
    \centering
    \includegraphics[width=0.5\linewidth]{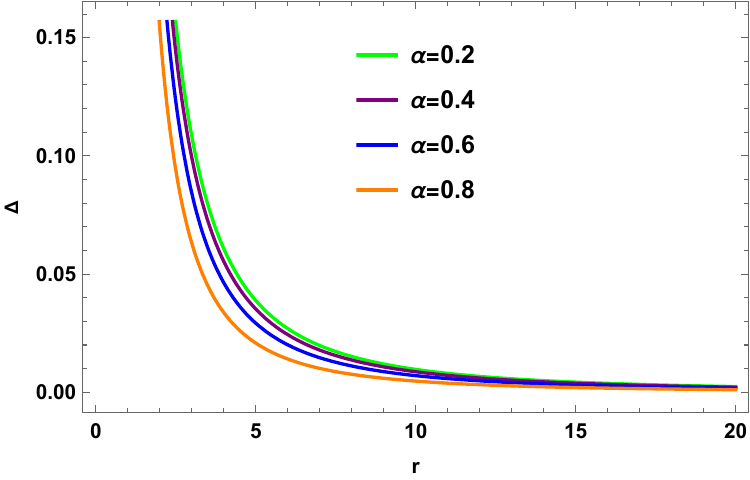}
    \caption{The nature of anisotropy parameter $\Delta$ with $r$ for the shape function $A(r)=r_0\,\left(r/r_0\right)^n$ with $n=1/2<1$, throat radius $r_0=0.2$.}
    \label{fig:28}
\end{figure}

We have plotted this ansiotropy parameter $\Delta$ in Figure \ref{fig:28} with the radial distance $r$ keeping fixed the throat radius $r_0=0.2$ and different values of the global monopole parameter $\alpha$. We see that this parameter remains positive for $r>r_0$ indicating that the wormhole solution is repulsive in nature.

Thus, we conclude that the wormhole model with the global monopole charge for the chosen shape function $A(r)=r_0\,\left(r/r_0\right)^n$ is an example of positive energy-density traversable wormhole solution in general relativity satisfying partially the energy conditions.

\section{Conclusions}

Wormholes are fascinating objects in theoretical physics that connect two distinct regions of spacetime. In general relativity (GR), traversable wormholes require the violation of the energy conditions (ECs). Finding exact wormhole solutions in GR, particularly with a known equation of state (EoS), is a complex task. Although there is currently no observational evidence confirming the existence of wormholes, they are predicted by various gravitational theories, much like black holes. Studying wormholes is crucial, as understanding their properties could aid in their future detection.

One of the key challenges in wormhole research is discovering exact solutions while minimizing the amount of exotic matter required. Despite the lack of experimental detection, this study have explored the potential existence of some wormhole geometries, considering global monopole charge within the framework of GR and an anisotropic fluid as the matter-energy context. We then analyzed different energy conditions, such as the weak energy, strong energy, and dominant energy conditions, for the chosen shape functions discussed in subsection \ref{subsec:1}--\ref{subsec:7}. Notably, the models presented in subsections \ref{subsec:3}, \ref{subsec:5}, and \ref{subsec:7} provided examples of topologically charged wormhole solutions in general relativity without requiring exotic matter. We demonstrated that the global monopole charge, characterized by the parameter $\alpha$, influences the energy conditions. However, other models presented in subsections \ref{subsec:1}, \ref{subsec:2}, \ref{subsec:4}, and \ref{subsec:6} also represent topologically charged wormhole solutions without exotic matter, although one or more energy conditions are partially violated. Additionally, we examined the anisotropy parameter in all chosen models in subsections \ref{subsec:1}--\ref{subsec:7} and showed that the presence of global monopole parameter $\alpha$ in wormhole space-times determines whether the wormhole geometry is repulsive or attractive in nature. The energy conditions for model-{\bf VII} are listed in Tables \ref{table:1}--\ref{table:3} and models-{\bf I} to {\bf VI} are listed in Table \ref{table:5}.

The comprehensive study of Morris-Thorne-type wormholes with global monopole charge and the examination of energy conditions associated with various shape functions enhance our understanding of their potential existence in the universe. While the speculation about the necessity of exotic matter with negative energy to sustain wormholes remains though a key point, the pursuit of such elusive elements and the fascination with the prospect of space-time shortcuts continues to captivate both physicists and science fiction enthusiasts. This work, therefore, contributes significantly to the ongoing discourse on the theoretical possibilities and implications of these enigmatic passages in the fabric of space-time, especially in the presence of global monopole charge.

In our future studies, we aim to study the stability using different approaches as well as linear perturbation of this topologically charged Morris-Thorne-type wormhole, providing a detail analysis of the results. Specifically, we will investigate gravitational perturbations of the wormhole using the Newman-Penrose formalism and calculate quasinormal modes (QNMs). Additionally, we will explore related topics such as wormhole evolution, quantum effects in wormhole space-times, astrophysical signatures of wormholes, and gravitational lensing.

\section*{Acknowledgement}

We sincerely acknowledges the anonymous referees for their valuable comments and helpful suggestions. F.A. acknowledges the Inter University Centre for Astronomy and Astrophysics (IUCAA), Pune, India for granted visiting associateship.

\section*{Data Availability Statement}

This manuscript has no associated data.

\section*{Code Availability Statement}

This manuscript has no associated code/software.

\end{document}